\documentclass[12pt]{iopart}
\usepackage{iopams}
\usepackage{epsfig}

\begin{document}

\tolerance = 10000


\title{Charge and Spin Transport in the One-dimensional Hubbard Model}
\author{Shi-Jian Gu$^{1,2}$, N. M. R. Peres$^3$, J. M. P. Carmelo$^4$}
\address{$^1$Department of Physics and Institute of Theoretical Physics, The
Chinese University of Hong Kong, Hong Kong, China\\$^2$Zhejiang Institute of
Modern Phys,
Zhejiang University, Hangzhou, 310027, China\\
$^3$Center of Physics and Physics Department of the University of Minho, School
of Sciences, Campus of Gualtar, P-4710-057 Braga, Portugal\\ $^4$ GCEP-Center
of Physics, University
of Minho, Campus Gualtar, P-4710-057 Braga, Portugal\\
E-mail: sjgu@phy.cuhk.edu.hk}

\begin{abstract}
In this paper we study the charge and spin currents transported by the
elementary excitations of the one-dimensional Hubbard model. The corresponding
current spectra are obtained by both analytic methods and numerical solution of
the Bethe-ansatz equations. For the case of half-filling, we find that the
spin-triplet excitations carry spin but no charge, while charge $\eta$-spin
triplet excitations carry charge but no spin, and both spin-singlet and charge
$\eta$-spin-singlet excitations carry neither spin nor charge currents.
\end{abstract}


\maketitle

\section{Introduction}

Recently, there has been a renewed interest in the unusual transport and
spectral properties of nanotubes, ballistic wires, and quasi-one-dimensional
(1D) compounds \cite{Hiro,Sing}. Quantum effects are strongest at low
dimensionality leading to unusual phenomena such as charge-spin separation at
all energies \cite{Sing} and persistent currents in mesoscopic rings
\cite{Eckern}. Thus, the further understanding of the transport of charge in
low-dimensional  correlated systems and materials is a topic of high scientific
interest.

There is numerical evidence of a fundamental difference between the transport
properties of integrable and nonintegrable 1D interacting quantum systems: at
finite temperatures, $T>0$, the integrable systems behave as ideal conductors in
the metallic quantum phases and as ideal insulators in the insulating phases,
with the concepts of an ideal insulator and conductor defined in Ref.
\cite{Prelovsek}. In contrast, the nonintegrable 1D interacting systems are
generic conductors and activated ones in the metallic and insulating phases,
respectively. While in the trivial case of 1D integrable systems whose
Hamiltonians commute with the current operator the ideal insulating and
conducting behaviors are easy to confirm, there is the expectation that such
ideal behaviors might also occur in 1D integrable quantum systems whose
Hamiltonian does not commute with that operator, such as the 1D Hubbard model
\cite{JHubbard63,EHLieb68,Takahashi}. However, the studies of Ref. \cite{Kawa}
rely on the generalization of the thermodynamic Bethe-ansatz equations introduced
in Ref. \cite{Takahashi} to the model in a presence of a vector potential \cite{Nuno97}
and seem to reveal that for half filling the 1D Hubbard model does not remain an
ideal insulator for $T>0$, in contrast to the general predictions and
expectations of Ref. \cite{Prelovsek}.

Solvable lattice models such as the 1D Hubbard model
\cite{JHubbard63,EHLieb68,Takahashi,YQLi98} and the XXZ chain \cite{HABethe31}
are often used as toy effective models for the study of the unusual properties
of quasi-1D compounds \cite{poly87,spectral0}. Although the 1D Hubbard model was
diagonalized long ago \cite{EHLieb68,Takahashi} by means of the coordinate
Bethe ansatz (BA) \cite{CNYang67,MGaudin67}, the involved form of BA wave
functions has prevented the full calculation of dynamic response functions. The
study of the asymptotic of correlation functions and of the low-energy
dynamical properties was performed by combining the BA solution with other
methods, such as conformal-field theory \cite{HFrahm9091}, bosonization
\cite{HJSchulz90,boso94}, the pseudo-particle formalism\cite{JMPCarmelo92}, and
scaling methods \cite{CAStafford93}.

In this paper, we study the charge and spin currents carried by the elementary low-energy
and finite-energy excitations of the 1D Hubbard model. The paper is organized as follows:
In Sec. II we summarize the basic information about the model and the BA solution needed
for our study; The energy and current spectra of the elementary excitations is the
subject of Sec. III. Finally, in Sec. IV we present the concluding remarks.

\section{The model and its Bethe-ansatz solution}\label{sec_hub:model}

The Hamiltonian of the 1D Hubbard model on a periodic $N_a$-site chain reads,
\begin{equation}
{\cal H}= -t\sum_{j,\, \sigma}(c_{j,\,\sigma}^\dagger
c_{j+1,\,\sigma}+ c_{j+1,\,\sigma}^\dagger c_{j,\,\sigma}) + U
\sum_j (n_{j,\uparrow}-1/2)(n_{j,\downarrow}-1/2) \, ,
\label{hub_eq:Hamiltonian1}
\end{equation}
where the operator $c_{j,\,\sigma}^\dagger$ (and $c_{j,\,\sigma}$) creates (and
annihilates) an electron of spin projection $\sigma$ at the site of index
$j=1,2,3,...,N_a$ and $n_{j,\,\sigma}=c_{j,\,\sigma}^\dagger c_{j,\,\sigma}$ is the
number operator at the same site. We use units of lattice constant one such that $L=N_a$,
where $L$ is the system length. We denote the electron number and the spin-projection
$\sigma$ electron number by $N$ and $N_{\sigma}$, respectively, such that
$N=[N_{\uparrow}+N_{\downarrow}]$. Moreover, we denote the states spin and $\eta$-spin
values by $S$ and $\eta$, respectively.

The model as written in Eq. (\ref{hub_eq:Hamiltonian1}) has both a spin and a
$\eta$-spin $SU(2)$ symmetry \cite{LiebB,hub_CNYang89,hub_CNYang90,TDeguchi00}.
The generators of the $\eta$-spin symmetry are given by,
\begin{eqnarray}
&&\eta=\sum_{j=1}^L (-1)^j c_{j\uparrow}c_{j\downarrow},\;\;\;
\eta^\dagger=\sum_{j=1}^L (-1)^j c_{j\downarrow}^\dagger
c_{j\uparrow}^\dagger, \nonumber \\
&&\eta^z=\frac{1}{2}\sum_{j=1}^L
(n_{j\downarrow} + n_{j\uparrow})-\frac{1}{2}L, \nonumber \\
&&[\eta, \eta^\dagger]=-2\eta^z,\;\;[\eta, \eta^z]=\eta,\;\;\;[\eta^\dagger,
\eta^z]=-\eta^\dagger \, . \label{hub_eq:SO4generators}
\end{eqnarray}
The global symmetry of the model (\ref{hub_eq:Hamiltonian1}) corresponding to these
two $SU(2)$ symmetries is $SO(4)$, since
half of the irreducible representations of $SU(2)\bigotimes SU(2)$ are excluded. The BA
solution refers to the Hilbert subspace spanned by the lowest-weight states (LWSs) of
both the spin and $\eta$-spin algebras. The subspace which is not associated with such a
solution is spanned by the energy eigenstates obtained by applying onto the LWSs one of
the off-diagonal generators of the corresponding two algebras \cite{hub_Essler92}.

The charge and spin current operators of the model read \cite{JMPCarmelo92},
\begin{eqnarray}
&&{J}^\rho=-eit\sum_{\sigma}\sum_{j=1}^L(c_{j\sigma}^\dagger
c_{j+1\sigma}-
c_{j+1\sigma}^\dagger c_{j\sigma} ) \nonumber \\
&&{J}^{\sigma_z}=-\frac{1}{2}it\sum_{\sigma}\sum_{j=1}^L \sigma (c_{j\sigma}^\dagger
c_{j+1\sigma}- c_{j+1\sigma}^\dagger c_{j\sigma} ) \, . \label{hub_eq:CurrentOperator}
\end{eqnarray}

In order to calculate the expectation values of the charge and spin current operators it
is convenient to consider a uniform vector potential $A_x\,{\vec{e}}_x$, which modifies the
hopping term along the chain by the usual Peierls phase factor, $t\rightarrow t \exp(\pm
i\phi_\sigma/L)$. Following such a procedure, the Hamiltonian becomes,
\begin{eqnarray}
{\cal H} &=& -t\sum_{j, \sigma}(c_{j\sigma}^\dagger
c_{j+1\sigma}e^{i\phi_\sigma /L}+
 c_{j+1\sigma}^\dagger c_{j\sigma} e^{-i\phi_\sigma /L})\nonumber  \\ && + U \sum_j
(n_{j,\uparrow}-1/2)(n_{j,\downarrow}-1/2) \, .
\label{hub_eq:Hamiltonian2}
\end{eqnarray}

For a given energy eigenstate $|m \rangle$, the charge and spin current expectation
values $J_m^\rho = \langle m|J^\rho|m \rangle$ and $J_m^{\sigma_z} = \langle
m|J^{\sigma_z}|m \rangle$, respectively, can be expressed as follows \cite{BSShastry90},
\begin{eqnarray}
&&J_{m}^\rho=\left. \frac{d
E_m(\phi)}{d(\phi/L)}\right|_{\phi=0}\;\;\;\;\;\;\;\phi=\phi_\uparrow=\phi_\downarrow,
\nonumber
\\
&&J_{m}^{\sigma_z}=\left.\frac{d E_m (\phi)
}{d(\phi/L)}\right|_{\phi=0}\;\;\;\;\;\;\phi=\phi_\uparrow=-\phi_\downarrow.
\label{hub_eq:jsjqdef}
\end{eqnarray}

The $\phi_\sigma>0$ Hamiltonian (\ref{hub_eq:Hamiltonian2}) remains integrable and can be
diagonalized by means of coordinate BA \cite{Nuno97}. One can introduce two generalized $SU(2)$
symmetries for the $\phi_\sigma>0$ case \cite{new}. Since $2\eta$ and $S$ remain good
quantum numbers, one finds that the BA solution refers to the LWSs of both the
$\eta$-spin and spin generalized algebras. Thus, all the energies of the tower of states
such that $2\eta > [L -N]$ and $2S>[N_{\uparrow}-N_{\downarrow}]$ have the same energy as
the corresponding LWSs. It follows that the BA numbers can be related to the values
$\eta$ and $S$ of the states of each $\eta$-spin and spin tower, respectively. Since the
studies of the ensuing section refers to both LWSs and non-LWSs, here we express the sum
rules of the BA numbers in terms of the good quantum numbers $\eta$ and $S$. Moreover, we
provide the simplified expressions in terms of the electronic numbers $N$ and
$N_{\sigma}$ which correspond to the LWS of each tower only.

The solution of the Hamiltonian (\ref{hub_eq:Hamiltonian2}) by the BA leads to the
following equations \cite{Nuno97,Nuno},
\begin{eqnarray}
&&e^{ik_j L}=e^{i\phi_\uparrow}\prod_{\beta=1}^M\frac{\sin
k_j-\Lambda_\beta+iu}
{\sin k_j-\Lambda_\beta-iu}, \nonumber \\
&&e^{i(\phi_\downarrow-\phi_\uparrow)}\prod_{j=1}^{N_c}\frac{\Lambda_\gamma-\sin
k_j+iu}{\Lambda_\gamma-\sin k_j-iu}\nonumber \\
&&=-\prod_{\beta=1}^M\frac{\Lambda_\gamma-\Lambda_\beta+i2u}
{\Lambda_\gamma-\Lambda_\beta-i2u} \, . \label{hub_eq:BAEOriginal}
\end{eqnarray}
Here and throughout this paper $u=U/4t$, the numbers $N_c$ and $M$ such that $0\leq
N_c\leq N$ and $0\leq M\leq N_{\downarrow}$, respectively, are defined below, and
$\Lambda$ is the spin rapidity \cite{Takahashi}. Takahashi string hypothesis states that
in addition to the real solutions for $\Lambda_\gamma$ and $k_j$, there are solutions
involving complex $k_j$ and $\Lambda_\gamma$ values. The spin string $\Lambda$s of legth
$n$ is characterized by \cite{Takahashi},
\begin{eqnarray}
\Lambda_\gamma^{nj}=\Lambda_\gamma^n + (n+1-2j)iu.\;\;\;\;\; j=1,2, \dots, n. \, ,
\label{hub_eq:LambdaDef}
\end{eqnarray}
where $\Lambda_\gamma^n$ is the real part of the complex number. The charge $k-\Lambda$
string of length $n$ includes $2n$ $k$s and $n$ $\Lambda$s such that,

\begin{eqnarray}
&&{\Lambda '}_\gamma^{nj}={\Lambda '}_\gamma^{n}+
(n+1-2j)iu,\;\;\;\; n=1,2,\dots, n, \nonumber \\
&&k_\gamma^1=\pi -\sin^{-1}({\Lambda '}_\gamma^{n}+niu), \nonumber
\\
&&k_\gamma^2=\sin^{-1}({\Lambda '}_\gamma^{n}+(n-2)iu), \nonumber
\\
&&k_\gamma^3=\pi-k_\gamma^2,\nonumber \\
&& k_\gamma^4=\sin^{-1}({\Lambda '}_\gamma^{n}+(n-4)iu), \nonumber
\\
&& k_\gamma^5=\pi-k_\gamma^4, \nonumber \\
&&\dots,\nonumber \\
&&k_\gamma^{2n-2}=\sin^{-1}({\Lambda
'}_\gamma^{n}-(n-2)iu),\nonumber \\
&& k_\gamma^{2n-1}=\pi-k_\gamma^{2n-2},\nonumber \\
&&k_\gamma^{2n}=\pi - \sin^{-1}({\Lambda '}_\gamma^{n}-niu).
\label{hub_eq:KLambdaDef}
\end{eqnarray}
By use of Eqs. (\ref{hub_eq:LambdaDef}) and (\ref{hub_eq:KLambdaDef}) in Eqs.
(\ref{hub_eq:BAEOriginal}) we arrive to the following transcendental equations
\cite{Nuno,NMRPeres01},
\begin{eqnarray}
k_j L=2\pi I_j + \phi_\uparrow
-\sum_{n=1}^\infty\sum_{\beta=1}^{M_n}\theta(\frac{\sin k_j
-\Lambda_\beta^n }{nu})
-\sum_{n=1}^\infty\sum_{\beta=1}^{M'_n}\theta(\frac{\sin k_j
-{\Lambda '}_\beta^n }{nu}) \nonumber \\
L(\sin^{-1}({\Lambda '}_\gamma^n+inu)+\sin^{-1}({\Lambda
'}_\gamma^n- inu ))=2\pi {J'}_\gamma^n -n(\phi_\uparrow+\phi_\downarrow)\nonumber \\
+\sum_{j=1}^{N-2M'} \theta(\frac{{\Lambda '}_\gamma^n -\sin
k_j}{nu}) + \sum_{m, \beta}\Theta_{nm}(\frac{{\Lambda
'}_\gamma^n-{\Lambda
'}_\beta^m}{u}), \nonumber \\
\sum_{j=1}^{N-2M'}\theta(\frac{{\Lambda}_\gamma^n-\sin k_j}{nu}) = 2\pi J_\gamma^n
+n(\phi_\downarrow-\phi_\uparrow)+ \sum_{m,\beta}\Theta_{nm}(\frac{{\Lambda
}_\gamma^n-{\Lambda}_\beta^m}{u}) \, , \label{hub_eq:BAEEE}
\end{eqnarray}
where $\theta(x)=-2\tan(x)$ and
\begin{eqnarray}
\Theta_{nm}(x)&=&\theta\left(\frac{x}{|n-m|}\right)+2
\theta\left(\frac{x}{|n-m|+2}\right) + \cdots +2
\theta\left(\frac{x}{n+m-2}\right) \nonumber \\&&
 +\theta\left(\frac{x}{n+m}\right) \, , \;\;\;
{\rm for:} \; n\neq m \nonumber \\
&=& 2\theta\left(\frac{x}{2}\right)+2 \theta\left(\frac{x}{4}\right)+\cdots + 2
\theta\left(\frac{x}{2n-2}\right) +\theta\left(\frac{x}{2n}\right) \, , \nonumber \\ && {\rm
for: }\; n=m \, .
\end{eqnarray}
Here $\{I_j, {J'}_\gamma^n, J_\gamma^n\}$ are the actual quantum numbers whose
values define the energy eigenstates and thus determine the energy and current
spectra of the elementary excitations studied in the ensuing section. Following
the notation of Takahashi \cite{Takahashi}, we introduce the numbers,
\begin{eqnarray}
M_c=\sum_n n M'_n,\;\;\;\; M_s=\sum_n n M_n \, .
\end{eqnarray}
In these expressions $M_n$ and $M'_n$ are the numbers of spin $\Lambda$ strings of length
$n$ and charge $k-\Lambda$ strings of length $n$, respectively. The values of $N_c$ and
$M$ are then uniquely defined by the following sum rules,
\begin{eqnarray}
N_c =L -2\eta -2 M_c,\;\;\;\;\;  M =M_c+M_s = {L\over 2} - \eta - S \, ,
\end{eqnarray}
which for a LWS of the $\eta$-spin and algebras such that $2\eta = [L -N]$ and $2S
=[N_{\uparrow}-N_{\downarrow}]$, respectively, simplify to,
\begin{eqnarray}
N_c =N -2 M_c,\;\;\;\;\;  M = N_{\downarrow} \, .
\end{eqnarray}
The above quantum numbers $\{I_j, {J'}_\gamma^n, J_\gamma^n\}$ can be integers or
half-odd integers (HOI) according to the following prescriptions: $I_j$ is integer (or
HOI) if $\sum_m (M_m+M'_m)$ is even (odd); $J_\gamma^n$ is integer (HOI) if $N_c-M_n$ is
odd (even); ${J'}_\gamma^n$ is integer (HOI) if $L-(N_c-M'_n)$ is odd (even). They have
values in the following ranges,
\begin{eqnarray}
&& I_j<\frac{L}{2},\nonumber \\
&& {J'}_\gamma^n<\frac{1}{2}(L-N+2M_c-\sum_{m=1}^\infty
t_{nm}M'_m), \nonumber \\
&& J_\gamma^n<\frac{1}{2}(N-2M_c-\sum_{m=1}^\infty t_{nm} M_m) \, ,
\end{eqnarray}
where $t_{nm}= 2\min (n, m)-\delta_{nm}$.

The energy and momentum spectra are given in terms of the BA quantum numbers as follows,
\begin{eqnarray}
E&=&-\sum_{j=1}^{N_c} 2t\cos k_j +\sum_{n,\alpha}4t {\mathcal{R}}
\sqrt{1-({\lambda^{\prime }}^n_\alpha -inu)^2} \nonumber \\&& - {U\over
2}\Bigl(N_c + 2M_c - {1\over
2}\Bigr)  \, ,\nonumber \\
P&=&\frac{2\pi}{L}\Bigl(\sum_j I_j+\sum_{n, \alpha}{J}_\alpha^n\Bigr) +
\sum_{n,\alpha}\Bigl(\pi-\frac{2\pi}{L}{J'}_\alpha^n\Bigr) \nonumber \\&& +
\pi\Bigl(M_c+\eta-{1\over 2}[L -N]\Bigr) \, , \label{EP}
\end{eqnarray}
where $\mathcal{R}$ refers to the real part and $\pi (M_c+\eta-{1\over 2}[L -N])$
simplifies to $\pi M_c$ for a LWS of the $\eta$-spin algebra.

All energy eigenstates associated with the BA solution are described by different
occupancy configurations of the quantum numbers appearing in the system of coupled
equations given in Eq. (\ref{hub_eq:BAEEE}). For example, for the ground state there is
no complex solution for Eqs. (\ref{hub_eq:BAEOriginal}) and $I_j$ and $J_\gamma$ are
successive numbers centered around zero. Thus, the quantum number occupancy configuration
for even $N_c=N=L$ and odd $N_c/2$ corresponds to,
\begin{eqnarray}
&&I_j=-\frac{N-1}{2}, -\frac{N-3}{2},\cdots,\frac{N-1}{2},\nonumber\\ &&
J_\gamma=-\frac{M-1}{2}, -\frac{M-3}{2},\cdots,\frac{M-1}{2} \, .
\label{eq:groundstateqn}
\end{eqnarray}

In some of the figures presented in the ensuing section we measure the energy relative to
the ground-state energy. Such a choice corresponds to the following general energy
spectrum,
\begin{eqnarray}
E_0 &=& -\sum_{j=1}^{N_c} 2t\cos k_j +\sum_{n,\alpha}4t {\mathcal{R}}
\sqrt{1-({\lambda^{\prime }}^n_\alpha -inu)^2} \nonumber \\&& - {U\over
2}\Bigl[N_c + 2M_c - {1\over 2}\Bigr] - 2\mu (N_a -N) - 2\mu_0 H
(N_{\uparrow}-N_{\downarrow}) \, , \label{E-GS}
\end{eqnarray}
where $\mu$ is the chemical potential, $\mu_0$ the Bohr magneton, and $H$ the magnetic
field.

Although the Bethe-ansatz equations (\ref{hub_eq:BAEEE}) refer to $L>>1$, we have used
these equations in the numerical study of finite-$L$ chains and obtained results for
several quantities in excellent numerical agreement with the known exact values. Thus, in
the ensuing section we use these equations to derive numerically the charge and spin
currents carried by the elementary excitations for finite values of $L$, as well as their
energy spectra. The obtained results are a good approximation for the corresponding
current spectra of the $L>>1$ quantum problem which the equations (\ref{hub_eq:BAEEE})
refer to.

\section{Current spectra for a finite-size system}
\subsection{The case of half-filling}
\label{hub_sec:four}

The zero-magnetization and half-filling ground state is both a spin and
$\eta$-spin singlet. Thus, the simplest elementary excitations are spin-triplet
excitations such that $(\eta =0,S=1)$, spin-singlet excitations such that
$(\eta =0,S=0)$ whose energy spectrum is degenerated to that of the
spin-triplet excitations, charge $\eta$-spin-triplet excitations such that
$(\eta =1,S=0)$, and charge $\eta$-spin-singlet excitations such that $(\eta
=0,S=0)$ whose energy spectrum is degenerated to that of the charge
$\eta$-spin-triplet excitations. The energy spectra of these four elementary
excitations have been previously studied \cite{MTakahashib}. For a comparison,
we evaluate the energy spectrum and the charge and spin currents of all the
energy eigenstates corresponding to the above four types of half-filling
elementary excitations. The energies considered in this subsection correspond
to the general energy spectrum provided in Eq. (\ref{EP}).

The half-filling and zero-magnetization ground state considered in this
subsection is characterized by charge and spin distributions given by,
\begin{eqnarray}
\rho_0(k)&=&\frac{1}{2\pi}+\frac{\cos k}{\pi}\int_0^{\infty}\frac{J_0(p)\cos (p\sin
k)}{1+e^{u|p|/2}}dp,\nonumber \\
\sigma_0(\lambda)&=&\frac{1}{8\pi u}\int_{-\pi}^\pi {\rm
sech}\left[\frac{2\pi}{u}(\lambda-\sin k)\right]dk \, ,
\end{eqnarray}
where $J_0$ is the Bessel function of zero order. Following standard BA procedures, the
evaluation of the energy, charge-current, and spin-current spectra studied below involves
the use of corresponding distributions for the elementary excitations.

{\it Spin-triplet excitations}. Such elementary excitations are obtained by introducing
``holes" in the spin distribution of the numbers {$J_\gamma$} relative to the
ground-state occupancy configuration. For each value of the excitation momentum and
energy there is a spin tower of three $S=1$ states, differing in the spin projections $0,
\pm 1$, but all having the same values,
\begin{eqnarray}
N_c=N=L, \;\;\;\;\ M=L/2-1 \, ,
\end{eqnarray}
and
\begin{eqnarray}
&&I_j=-\frac{N-2}{2}, -\frac{N-4}{2},\cdots,\frac{N}{2},\nonumber
\\ &&J_\gamma^1=-\frac{M+1}{2}, -\frac{M-3}{2},\cdots,\frac{M+1}{2} \, ,
\label{hub_eq:spintriplet}
\end{eqnarray}
for the BA numbers. Hence, there are two holes $\lambda_1^h$ and $\lambda_2^h$ in the
spin distribution. The BA equations become,
\begin{eqnarray}
&&k_j L= 2\pi I_j+\phi_\uparrow-2\sum_{\beta=1}^{L/2-1}
\tan^{-1}\frac{\sin k_j -\lambda_\beta}{u}\nonumber \\
&&\sum_{j=1}^N 2\tan^{-1}\frac{\lambda_\gamma-\sin k_j}{u}= 2\pi
J_\gamma+(\phi_\downarrow -\phi_\uparrow) \nonumber \\&&
+2\sum_{\beta=1}^{L/2-1}\tan^{-1}\frac{\lambda_\gamma-\lambda_\beta}{2u} \, .
\end{eqnarray}
Following the usual procedures of BA \cite{Takahashi,Nuno}, one then introduces the
following charge and spin distributions,
\begin{eqnarray}
\rho(k)=\frac{1}{2\pi}&+&\cos k \int K_1(\sin
k-\lambda)\sigma(\lambda)d\lambda,\nonumber \\
\sigma(\lambda)+\sigma^h(\lambda)&=&\int K_1(\lambda-\sin k)\rho(k)dk \nonumber \\
&&-\int K_2(\lambda-\lambda')\sigma(\lambda')d\lambda' \, ,
\label{eq:densityspintriplet}
\end{eqnarray}
where $\sigma^h(\lambda)=[\delta(\lambda-\lambda^h_1)+\delta(\lambda-\lambda_2^h)]/L$ and
$K_n(x)=nu/[\pi(n^2 u^2+x^2)]$. In the presence of the flux the momentum deviation
corresponding to this elementary excitation is given by,
\begin{eqnarray}
&&\triangle k_j L=\phi_\uparrow \nonumber \\ &&
\;\;\;-2\pi\sum_{\beta=1}^{L/2-1}K_{1}
(\sin k_j-\lambda_\beta)(\cos k_j \triangle k_j-\triangle \lambda_\beta),\nonumber \\
&&2\pi\sum_{j=1}^N K_1
(\lambda_\gamma-\sin k_j)(\triangle \lambda_\gamma- \cos k_j \triangle
k_j)=(\phi_\downarrow-\phi_\uparrow)\nonumber \\
&&\;\;\;+2\pi\sum_{\beta=1}^{L/2-1} K_2(\lambda_\gamma-\lambda_\beta)(\triangle
\lambda_\gamma-\triangle \lambda_\beta) \, .
\end{eqnarray}
Use of Eq. (\ref{eq:densityspintriplet}) then yields,
\begin{eqnarray}
&&\triangle k \rho(k)=\frac{\phi_\uparrow}{2\pi L}+\int K_1(\sin
k-\lambda) \triangle \lambda\sigma(\lambda)d\lambda,\nonumber \\
&& \triangle
\lambda[\sigma(\lambda)+\sigma^h(\lambda)]=\frac{\phi_\downarrow-\phi_\uparrow}{2\pi
L} -\int K_2(\lambda-\lambda')\Delta \lambda' \sigma(\lambda')d\lambda' \nonumber
\\&&\;\;\; +\int K_1(\lambda-\sin k)\cos k\triangle k \rho(k)dk \, .
\label{eq:deltaken}
\end{eqnarray}
The corresponding energy deviation is given by
\begin{eqnarray}
\triangle E(\phi)=2tL\int \sin k \rho(k) \triangle k dk \, . \label{eq:phienergydiff}
\end{eqnarray}
Our next task is the solution of the equations given in (\ref{eq:deltaken}). Inserting
the result obtained for $\Delta k \rho(k)$ into Eq. (\ref{eq:phienergydiff}) we find,
\begin{eqnarray}
\Delta E&=&  -\frac{t(\phi_\downarrow-\phi_\uparrow)}{8\pi u L}\int dk \sin k
\left[ \frac{{\rm sech}\left[\frac{\pi}{2u}(\sin
k-\lambda^h_1)\right]}{\sigma_0(\lambda_1^h)}\,-\, \frac{{\rm
sech}\left[\frac{\pi}{2u}(\sin k-\lambda^h_2)\right]}{\sigma_0(\lambda_2^h)}
\right],\nonumber \\ &&\label{eq:spintripenergyd}
\end{eqnarray}
where we have used the relation $\Delta \lambda\simeq
(\phi_\downarrow-\phi_\uparrow)/(4\pi L \sigma_0(\lambda)$, which was obtained
by Fourier transformation. It then follows that when
$\phi_\downarrow=\phi_\uparrow$ the charge current defined by Eq.
(\ref{hub_eq:jsjqdef}) vanishes.

\begin{figure*}
\includegraphics[width=15cm,height=7cm]{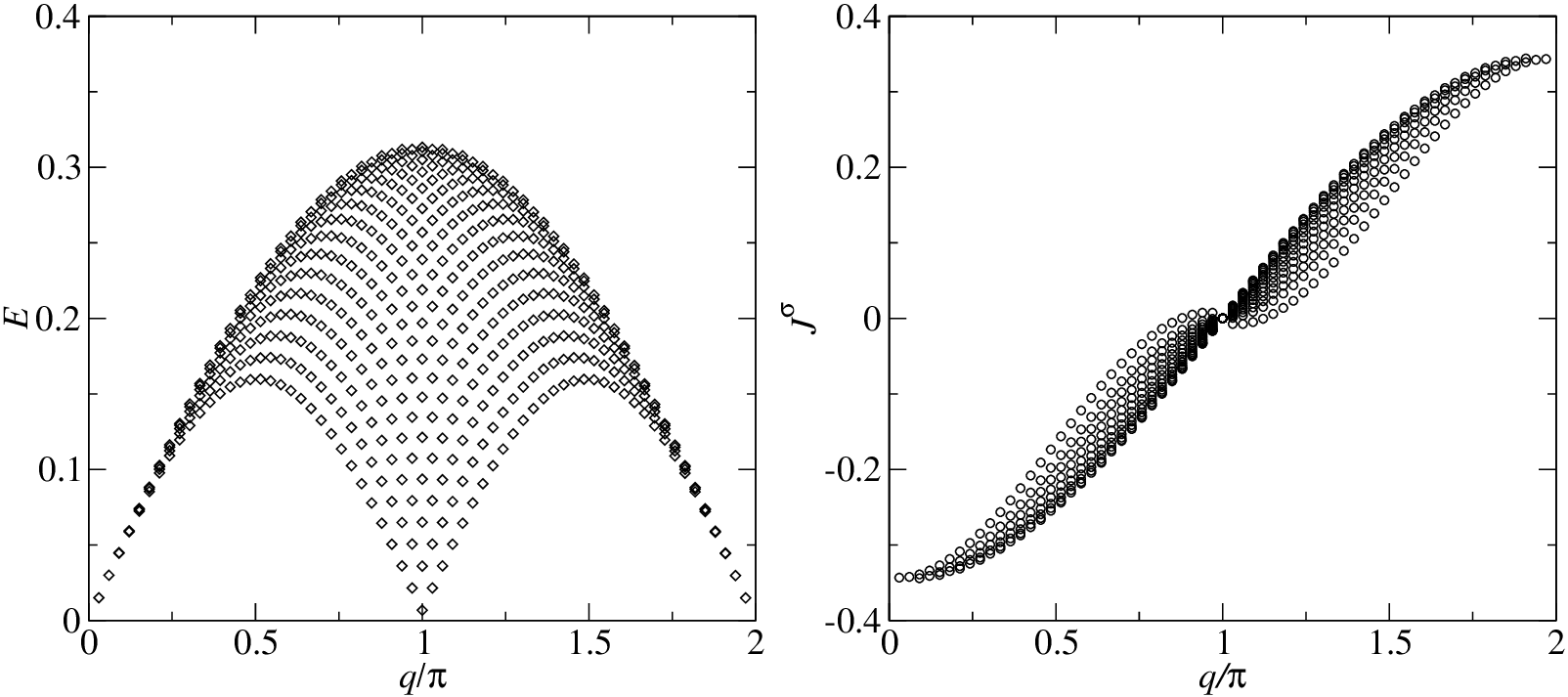}
\caption{\label{fig1} The half-filling energy spectrum ({\bf left}) and spin
current spectrum ({\bf right}) of the spin triplet excitations for $u=10$,
$N=L=66$, and $M=33$.\\}
\end{figure*}
\begin{figure*}
\includegraphics[width=15cm,height=7cm]{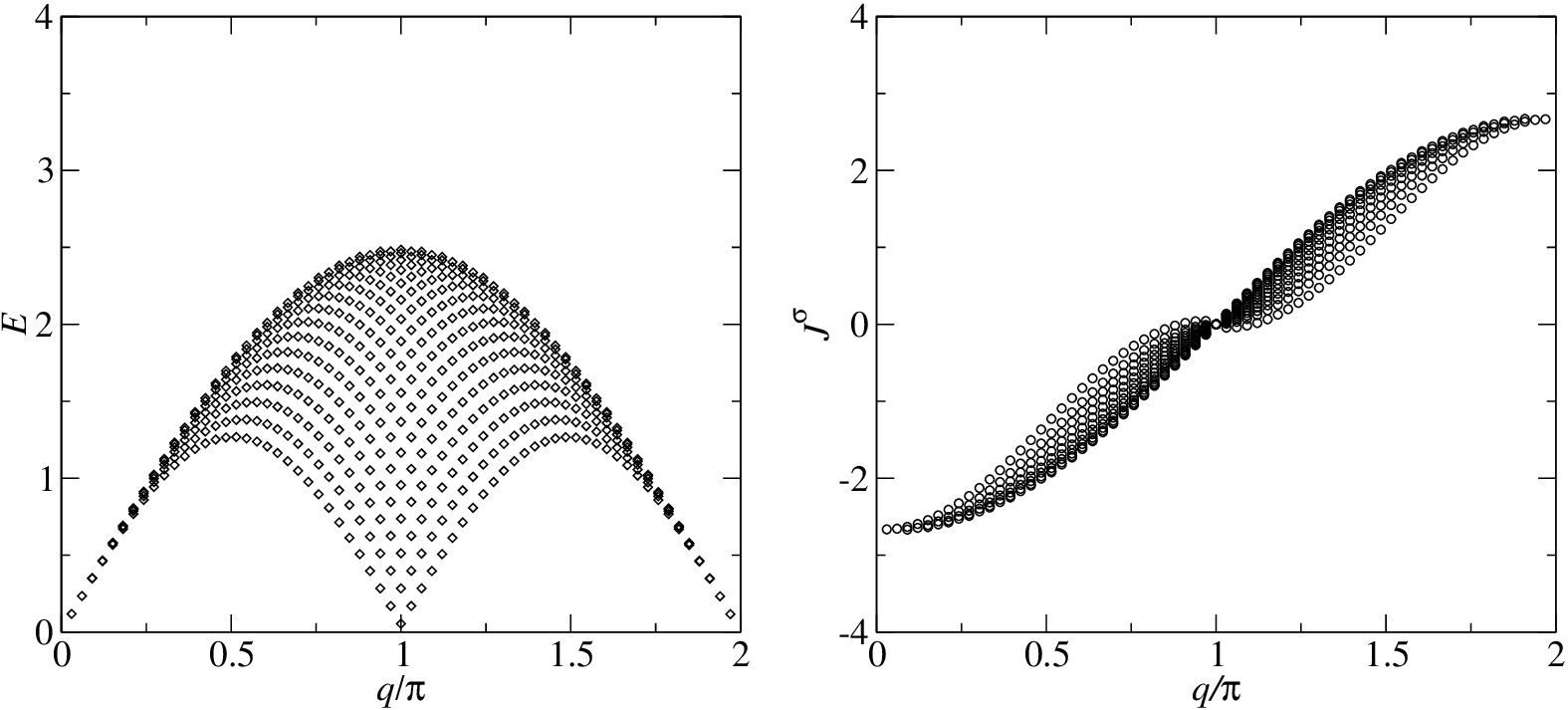}
\caption{\label{fig2} The half-filling energy spectrum ({\bf left}) and spin
current spectrum ({\bf right}) of the spin triplet excitations for $u=1$,
$N=L=66$, and $M=33$.\\}
\end{figure*}

In general case, $\lambda^h>1$, Eq. (\ref{eq:spintripenergyd}) can be solved
approximately with the result,
\begin{eqnarray}
J^\sigma\approx \frac{t\pi}{2u}\left[ \tanh\left(\frac{\pi \lambda^h_1}{2u}\right)
+\tanh\left(\frac{\pi \lambda^h_2}{2u}\right) \right] \, .
\end{eqnarray}
Since the momentum carried by the spin elementary excitation has the
form $q=2\tan^{-1}e^{-\lambda/2u}-\frac{\pi}{2}$ and the quantum
number of the charge part changes from half-integer to integer, what
generates a momentum shift $\pi$, the spin-current spectrum has the
following form, as was also observed in the antiferromagnetic
Heisenberg model \cite{QLZhang07},
\begin{eqnarray}
J^{\sigma}&=&\frac{t\pi}{2u}[\sin q_1+\sin q_2],
\nonumber \\
q&=&\pi+q_1+q_2,\;\;\;q_1, q_2\in[-\frac{\pi}{2}, \frac{\pi}{2}] \, .
\end{eqnarray}
Here $q_1$ and $q_2$ are the momenta of the two spin-distribution holes and $q$ denotes
the total excitation momentum. In figures 1 and 2 we plot the energy and spin-current
spectra of the spin-triplet excitations for $u=10$ and $u=1$, respectively. Such spectra
were obtained by solving numerically the BA equations.

The group velocity $v^{\sigma(\rho)}(q)$ and the effective spin (charge)
$e^{\sigma(\rho)}$ are defined as,
\begin{eqnarray}
v^{\sigma(\rho)}(q)=\frac{d E^{\sigma(\rho)}(q)}{dq},\;\;\;
e^{\sigma(\rho)}=e\frac{J^{\sigma(\rho)}(q)}{v^{\sigma(\rho)}(q)} \, .
\end{eqnarray}
In this equation $\sigma$ and $\rho$ denote the spin current and charge current,
respectively, and $e=-1$ and $e= 1/2$ for the charge and spin cases. Thus, the group
velocity of a single spin-distribution hole reads,
\begin{eqnarray}
v^\sigma(q)=-\frac{t\pi}{2u}\sin q \, ,
\end{eqnarray}
whereas the corresponding effective spin is given by,
\begin{eqnarray}
e^\sigma=-\frac{1}{2}.
\end{eqnarray}
Note that the total spin current is positive.

\begin{figure*}
\includegraphics[width=15cm,height=7cm]{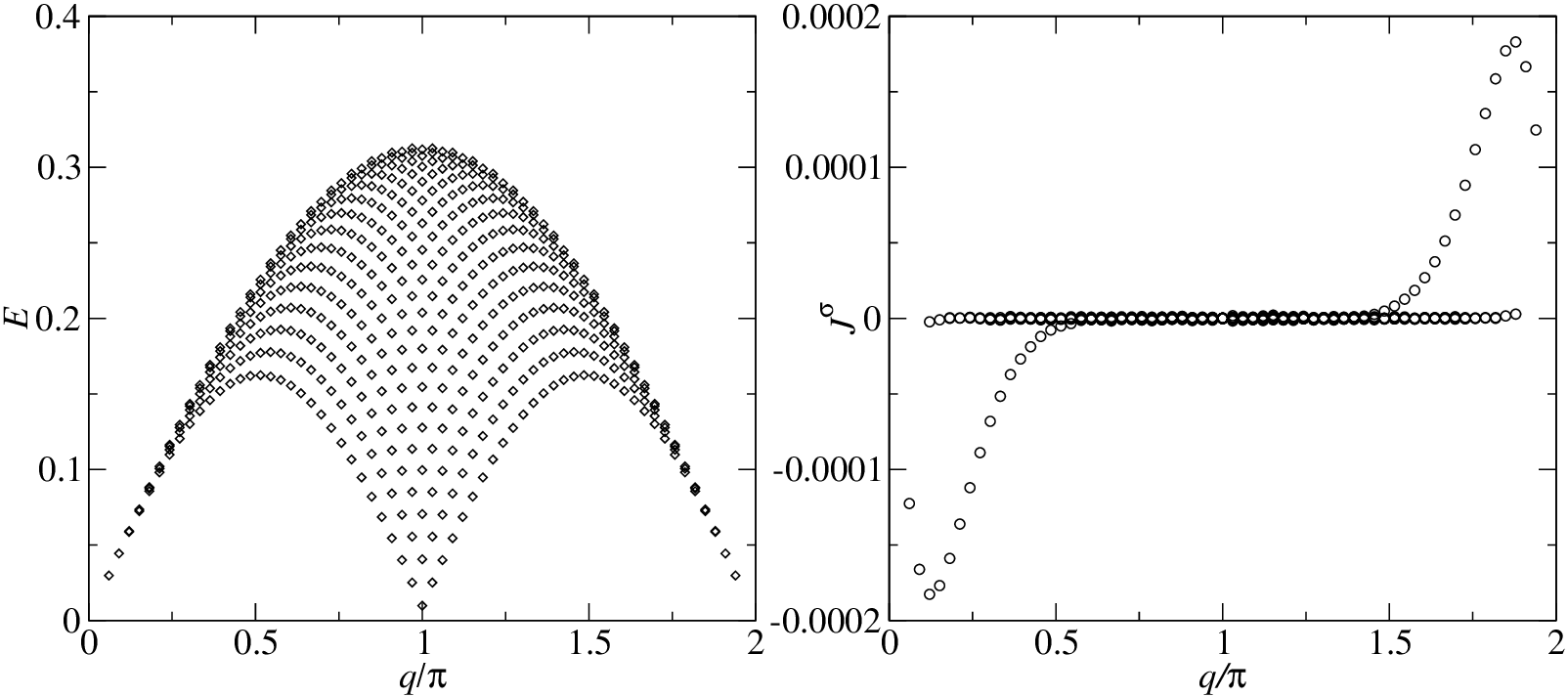}
\caption{\label{fig3} The half-filling energy spectrum ({\bf left}) and spin
current spectrum ({\bf right}) of the spin-singlet excitations for $u=10$,
$N=L=66$, and $M=33$.\\\\}
\end{figure*}
\begin{figure*}
\includegraphics[width=15cm,height=7cm]{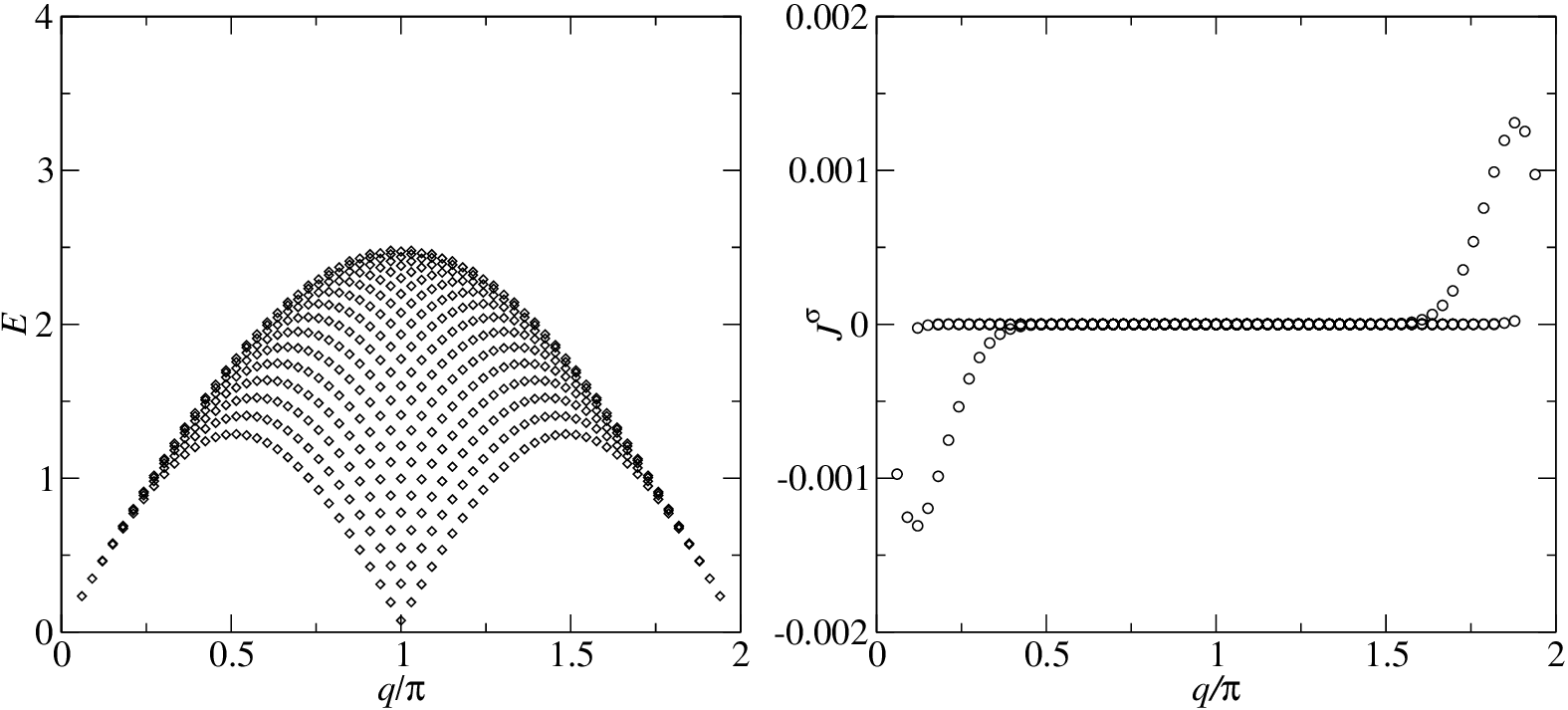}
\caption{\label{fig4} The half-filling energy spectrum ({\bf left}) and spin
current spectrum ({\bf right}) of the spin-singlet excitations for $u=1$,
$N=L=66$, and $M=33$.\\}
\end{figure*}

\begin{figure*}
\includegraphics[width=15cm,height=7cm]{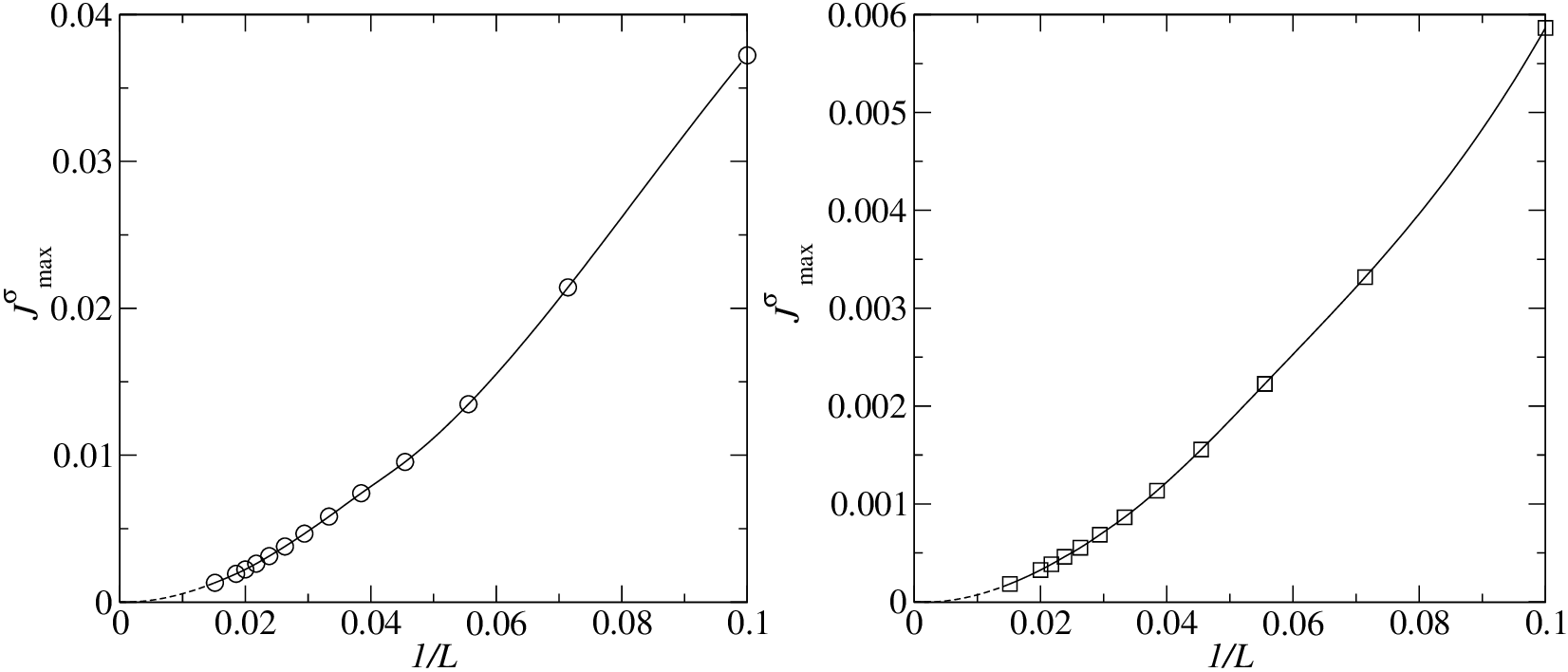}
\caption{\label{fig5} Scaling analysis of the maximum current carried by
spin-singlet excitations for $u=1$ (left) and $u=10$ (right) respectively.\\}
\end{figure*}

{\it Spin-singlet excitations}. The second type of elementary spin excitations studied
here corresponds to the spin-singlet excitations whose energy spectrum is degenerated
with that of the spin-triplet excitations considered above. Such excitations have a spin
string of length one. Thus, the BA numbers are given by,
\begin{eqnarray}
M_1=M-2,\;\;\; M_2=1 \, ,
\end{eqnarray}
and
\begin{eqnarray}
&&I_j=-\frac{N-2}{2}, -\frac{N-4}{2},\cdots,\frac{N}{2},\nonumber
\\ &&J_\gamma^1=-\frac{M-1}{2}, -\frac{M-3}{2},\cdots,\frac{M-1}{2}
\nonumber \\
&&J_0^2=0 \, . \label{hub_eq:spinsingletqnc}
\end{eqnarray}
It follows that there are again two holes in the spin distribution, $\lambda^h_1$ and
$\lambda^h_2$. The BA equations are given by,
\begin{eqnarray}
&&k_j L=2\pi I_j +\phi_\uparrow -2\sum_{\beta=1}^{L/2-2}
\tan^{-1}\frac{\sin k_j-\lambda_\beta}{u} \nonumber \\ && \;\;\;
-2\tan^{-1}\frac{\sin k_j-\Lambda}{2u},\nonumber \\ &&
2\sum_{j=1}^N\tan^{-1}\frac{\lambda_\gamma-\sin k_j}{u}=2\pi
J_\gamma +(\phi_\downarrow-\phi_\uparrow)\nonumber \\ && \;\;\;
+2\sum_{\beta=1}^{L/2-2}\tan^{-1}\frac{\lambda_\gamma-\lambda_\beta}{2u}
\nonumber \\
&&\;\;\;+2\tan^{-1}\frac{\lambda_\gamma-\Lambda}{u}
+2\tan^{-1}\frac{\lambda_\gamma-\Lambda}{3u}\nonumber \\
&& 2\sum_{j=1}^N\tan^{-1}\frac{\Lambda-\sin k_j}{2u} =2\pi
J^{(2)}_1(=0)+2(\phi_\downarrow-\phi_\uparrow) \nonumber \\ &&
\;\;\;+2\sum_{\beta}^{L/2-2}\left[\tan^{-1}\frac{\Lambda-\lambda_\beta}{u}+
\tan^{-1}\frac{\Lambda-\lambda_\beta}{3u}\right] \, ,
\label{eq:densityspinsinglet}
\end{eqnarray}
where $\Lambda$ denotes the rapidity of the spin string excitation of length two. The
deviations of the charge and spin distributions are such that,
\begin{eqnarray}
&&\triangle k \rho(k)=\frac{\phi_\uparrow}{2\pi L}+\int K_1(\sin
k-\lambda) \triangle \lambda\sigma(\lambda)d\lambda \nonumber \\ &&
\;\;\;+\frac{1}{L}K_2(\sin k-\Lambda)\Delta\Lambda \nonumber \\
&& \triangle
\lambda[\sigma(\lambda)+\sigma^h(\lambda)]=\frac{\phi_\downarrow-\phi_\uparrow}{2\pi
L}\nonumber \\
&&\;\;\; -\int K_2(\lambda-\lambda')\Delta\lambda'\sigma(\lambda')d\lambda'
\nonumber \\ &&\;\;\;
+\int\cos k\triangle k\rho(k) K_1(\lambda-\sin k)dk\nonumber \\ &&\;\;\;
-\frac{1}{L}[K_1(\lambda-\Lambda)+K_3(\lambda-\Lambda)]\Delta\Lambda \nonumber \\
&&\Delta \Lambda\sigma_2(\Lambda) =
\frac{\phi_\downarrow-\phi_\uparrow}{\pi L}
\nonumber \\ &&\;\;\;
-\int [K_1(\Lambda-\lambda)+K_3(\Lambda-\lambda)]\Delta \lambda
\sigma(\lambda)d\lambda \nonumber \\
&&\;\;\;+\int \cos k K_2(\Lambda-\sin k)\Delta k\rho(k) dk \nonumber \\ &&\;\;\; -\int
[2K_2(\Lambda-\Lambda')+K_4(\Lambda-\Lambda')]\Delta \Lambda \sigma_2 d\Lambda \, .
\label{eq:tttttdft}
\end{eqnarray}
Here $\sigma_2(\Lambda')=\delta(\Lambda'-\Lambda)/L$ and
$\sigma^h(\lambda)=[\delta(\lambda-\lambda_1^h)+\delta(\lambda-\lambda_1^h)]/L$.
Integrations in the variables $k, \lambda, \Lambda$ lead to,
\begin{eqnarray}
&&\int \cos k\Delta k\rho(k)=0 \nonumber \\
&&\int \Delta \lambda\sigma(\lambda)d\lambda +\frac{1}{L}\Delta\Lambda
=\int \frac{\phi_\downarrow-\phi_\uparrow}{4\pi L} d\lambda -
\int\Delta\lambda\sigma^h(\lambda)d\lambda,\nonumber \\
&&\frac{1}{L}\Delta\Lambda +\frac{1}{2}\int \Delta\lambda\sigma(\lambda) d\lambda =\int
\frac{\phi_\downarrow-\phi_\uparrow}{4\pi L} d\lambda \, .
\end{eqnarray}
Finally, the deviation associated with the spin string excitation of length two has the
form,
\begin{eqnarray}
\Delta\Lambda=(\Delta\lambda^h_1+\Delta\lambda^h_2) +\int
\frac{\phi_\downarrow-\phi_\uparrow}{4\pi} d\lambda \, ,
\end{eqnarray}
where the second term on the right-hand side can be omitted because it does not
contribute to the spin current. Use of this expression in the first equation of
(\ref{eq:tttttdft}) leads to,
\begin{eqnarray}
\triangle k \rho(k)\approx \frac{\phi_\uparrow}{2\pi L}+\int K_1(\sin k-\lambda)
\triangle \lambda[\sigma+\sigma^h]d\lambda \, .
\end{eqnarray}
Since $\sigma_0(\lambda)\approx \sigma(\lambda)+\sigma^h(\lambda)$, by means of the same
procedure already used for spin-triplet excitation, we find that to first order the
energy deviation induced by the external flux vanishes. Thus, both the charge and spin
currents carried by this type of elementary excitation vanish.

In figures 3 and 4 we plot the energy and spin current spectra of the
spin-singlet excitations for $u=10$ and $u=1$, respectively. The two small
features of the spin-current spectrum result from finite-size effects and
disapear in the thermodynamic limit, as shown in figure 5.

We emphasize that although the group velocity of the two spin-distribution holes of the
spin-singlet excitations is finite, the corresponding effective spin vanishes.

\begin{figure*}
\includegraphics[width=15cm,height=7cm]{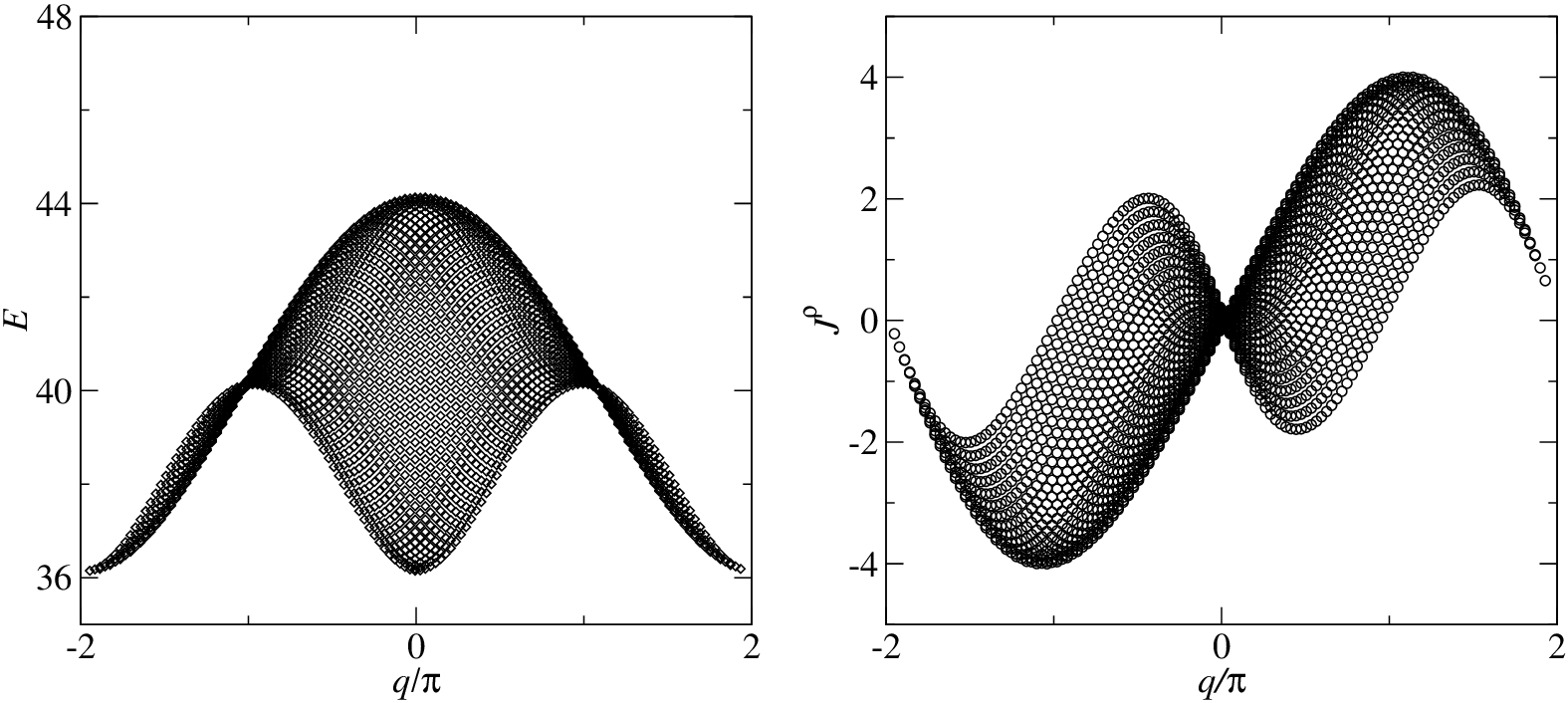}
\caption{\label{fig6} The half-filling energy spectrum ({\bf left}) and
charge-current spectrum ({\bf right}) of the charge $\eta$-spin triplet
excitation for $u=10$, $N=L=46$, and $M=23$.\\\\}
\end{figure*}
\begin{figure*}
\includegraphics[width=15cm,height=7cm]{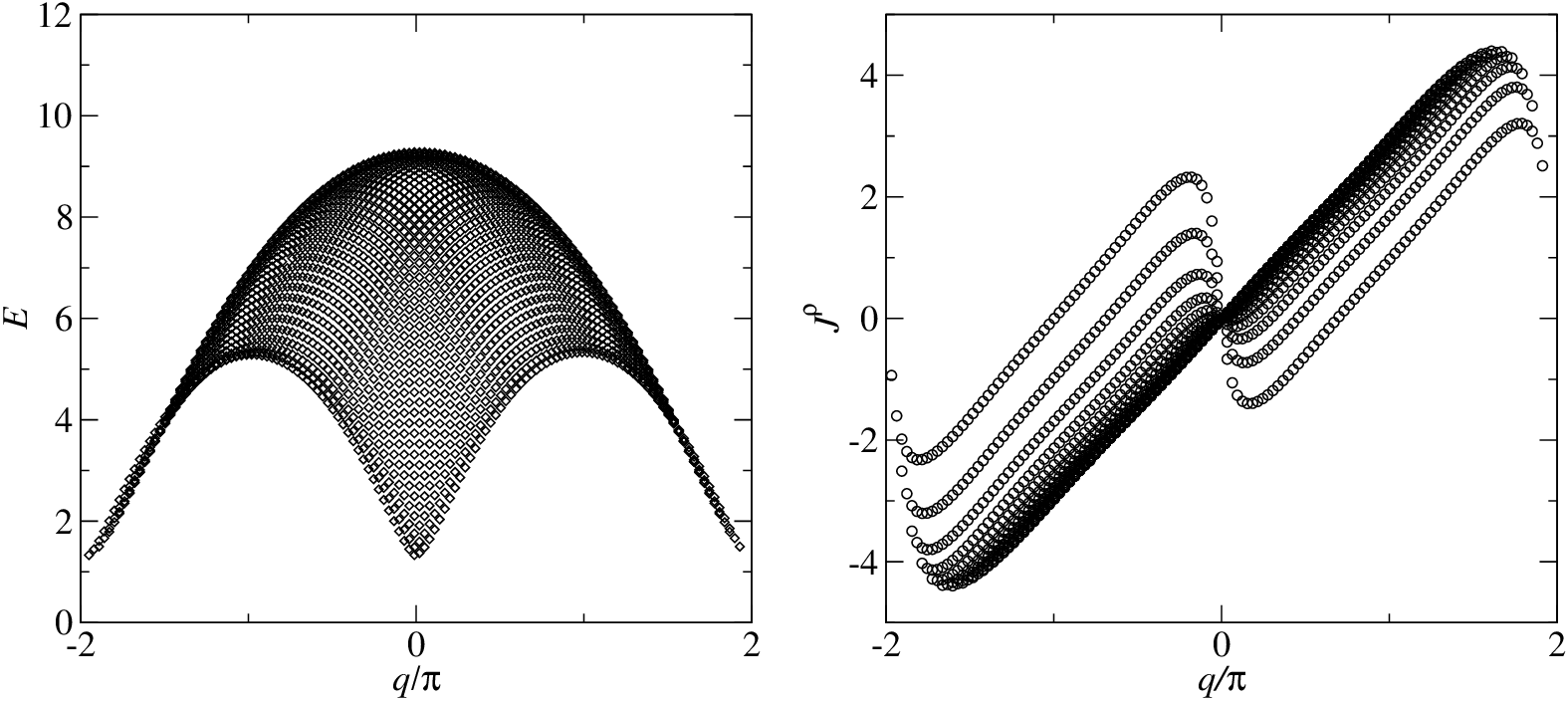}
\caption{\label{fig7} The half-filling energy spectrum ({\bf left}) and
charge-current spectrum ({\bf right}) of the charge $\eta$-spin triplet
excitation for $u=1$, $N=L=46$, and $M=23$.}
\end{figure*}

{\it Charge $\eta$-spin-triplet excitations}. For each value of the excitation momentum
and energy there are three types of such $\eta =1$ elementary excitations, which
correspond to the three values $0, \pm 1$ for the $\eta$-spin projection. All these
excitations have again the same BA numbers,
\begin{eqnarray}
N=L-2,\;\;\;\;\;\; M=N/2 \, ,
\end{eqnarray}
and
\begin{eqnarray}
&&I_j=-\frac{L}{2}, \cdots,\frac{L}{2}-1,\nonumber \\
&&J_\gamma=-\frac{M-1}{2},\cdots,\frac{M-1}{2} \, .
\end{eqnarray}
There are two holes $k^h_1$ and $k^h_2$ in the charge distribution such that,
\begin{eqnarray}
\rho^h(k)=\frac{1}{L} [\delta(k-k^h_1)+\delta(k-k^h_2)] \, .
\end{eqnarray}
Thus, the distributions $\rho(k)$ and $\sigma(\lambda)$ satisfy the following equations,
\begin{eqnarray}
&&\rho(k)+\rho^h(k)=\frac{1}{2\pi}+\cos k\int K_1(\sin k -\lambda)
  \sigma(\lambda) d\lambda \nonumber \\
&&\sigma(\lambda)=\int K_1(\lambda-\sin k)\rho(k) \nonumber \\
&&\;\;\;\;\;\;\;\;\;-\int K_2(\lambda-\lambda') \sigma(\lambda') d\lambda' \, .
\end{eqnarray}
The energy deviation corresponding to the two charge-distribution holes read,
\begin{eqnarray}
\Delta E(\phi)=-2t\left[\sin k_1^h \Delta k_1^h  + \sin k_2^h \Delta k_2^h \right] \, ,
\end{eqnarray}
where the momentum deviation $\Delta k$ is approximately given by,
\begin{eqnarray}
\Delta k \approx \frac{\phi_\uparrow}{2\pi L \rho_0(k)} \, .
\end{eqnarray}
It follows that the charge current spectrum is such that,
\begin{eqnarray}
J^\rho=-2t\left[\frac{\sin k^h_1}{\rho_0(k^h_1)} +\frac{\sin k^h_2}{\rho_0(k^h_2)}
\right] \, ,
\end{eqnarray}
while for $k^h=-k$ it reads,
\begin{eqnarray}
J^\rho&=&2t\left[\frac{\sin k_1}{\rho_0(k_1)} +\frac{\sin k_2}{\rho_0(k_2)}
\right], \nonumber \\
k&=& k_1+k_2 \, .
\end{eqnarray}
In turn, the spin current vanishes. In the strong coupling limit, $u=U/4t>>1$, one has
that $\rho_0(k)\simeq 1/2\pi$ and thus the spectrum simplifies to,
\begin{eqnarray}
J^\rho\propto \sin k_1+\sin k_2 \, .
\end{eqnarray}
In this case the corresponding energy spectrum can be expressed as the sum of three
cosine functions, in addition to the energy gap. Hence, the velocity of a single
charge-distribution hole simplifies to $v^\rho=-2t\sin k_h$, whereas the effective charge
is $-e$. This corresponds to a positive current in units of $e$.

Again, we used the BA equations to calculate the energy, charge-current, and
spin-current spectra for $u=10$ and $u=1$. (The spin current vanishes for the
charge $\eta$-spin-triplet states considered here.) The energy and
charge-current spectra are plotted in Figs. 6 and 7 for $u=10$ and $u=1$,
respectively. Note that the charge-current spectra features have a stronger
linear character for $u=1$ than for $u=10$. We interpret this effect as due to
the weak-coupling peak in the charge distribution as a function of the momentum
$k$. When a hole is created away from zero momentum it is less affected by the
other charges, leading to $J^\rho \propto q$.
\begin{figure*}
\includegraphics[width=15cm,height=7cm]{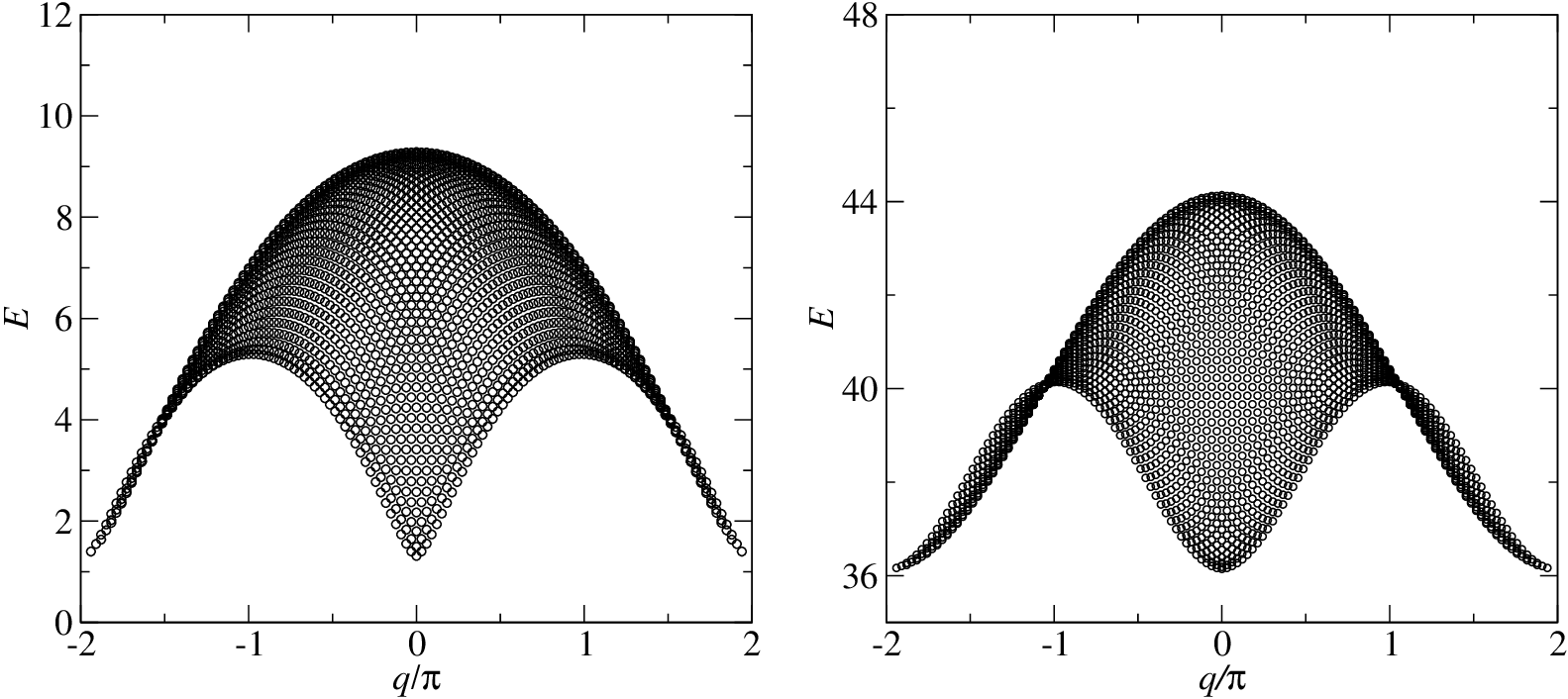}
\caption{\label{fig8} The half-filling energy spectra of charge $\eta$-spin
singlet excitation for the strong coupling $u=1$ ({\bf left}), and weak
coupling $u=10$ ({\bf right}).\\}
\end{figure*}

{\it Charge $\eta$-spin-singlet excitations}. These $\eta =0$ elementary excitations are
those whose energy spectrum is degenerated with that of the charge $\eta$-spin-triplet
excitations considered above. Such $\eta$-spin-singlet excitations contain one charge
string of length one. The BA numbers are then given by,
\begin{eqnarray}
M_1=\frac{N}{2}-1,\;\;\;M'_1=1 \, ,
\end{eqnarray}
and
\begin{eqnarray}
&&I_j=-\frac{N-1}{2},\cdots,\frac{N-1}{2},\nonumber \\
&&J_\gamma=-\frac{M_1-1}{2},
-\frac{M_1-3}{2},\cdots,\frac{M_1-1}{2}
\nonumber \\
&&J'_1=0 \, . \label{hub_eq:chargesinglet}
\end{eqnarray}
These excitations involve two holes $k^h_1$ and $k^h_2$ in the charge distribution. The
BA equations are such that,
\begin{eqnarray}
&&k_j L=2\pi I_j
+\phi_\uparrow-\sum_{\beta=1}^{L/2-1}2\tan^{-1}\frac{\sin k_j
-\lambda_\beta}{u}\nonumber \\ &&-2\tan^{-1}\frac{\sin k_j
-\Lambda}{u},
\nonumber \\ &&
\sum_{j=1}^{N-2}2\tan^{-1}\frac{\lambda_\gamma-\sin k_j}{u} = 2\pi
J_\gamma +(\phi_\downarrow-\phi_\uparrow)\nonumber
\\&& +\sum_{\beta=1}^{N/2-1}2\tan^{-1}\frac{\lambda_\gamma-\lambda_\beta}{2u},
\nonumber \\ && L[\sin^{-1}(\Lambda+i u)+\sin^{-1}(\Lambda-iu)] \nonumber \\ && =2\pi
{J'}_1^1-(\phi_\uparrow +\phi_\downarrow)+ 2\sum_{j=1}^{L-2}\tan^{-1}\frac{\Lambda-\sin
k_j}{u} \, ,
\end{eqnarray}
where $\Lambda$ is the rapidity involved in the charge string of length-one. Moreover, we
find,

\begin{eqnarray}
\Delta k_j(\rho+\rho^h)&=&\frac{\phi_\uparrow}{2\pi L}+\int
K_1(\sin k-\lambda)\Delta \lambda \sigma(\lambda)d\lambda \nonumber \\
&&+\frac{1}{L}K_1(\sin k-\Lambda)\Delta \Lambda,\nonumber \\
\Delta \lambda \sigma(\lambda)&=&\frac{\phi_\downarrow-\phi_\uparrow}{2\pi L}
+\int K_1(\lambda-\sin k)\cos k \Delta k dk\nonumber \\
&&-\int K_2(\lambda-\lambda')\Delta \lambda' \sigma(\lambda') d\lambda'.
\nonumber \\
\Delta\Lambda\sigma'(\Lambda)&=&-\frac{\phi_\downarrow+\phi_\uparrow}{2\pi L}
\nonumber \\
&&-\int K_1(\Lambda-\sin k)\cos k\Delta k\rho(k) dk \nonumber \\ && -\int
K_2(\Lambda-\Lambda')\sigma'(\Lambda')d\Lambda' \, ,
\end{eqnarray}
where $\sigma'(\Lambda')=\delta(\Lambda'-\Lambda)/L$. Integrations involving the
variables $\lambda, k$, and $\Lambda$ lead to,
\begin{eqnarray}
\Delta\Lambda=\frac{1}{2}[\cos k^h_1\Delta k^h_1+\cos k^h_2\Delta k^h_2] \, .
\end{eqnarray}
This is consistent with $\rho^h=[\delta(k-k_1^h)+\delta(k-k_2^h)]/L$ and $\Lambda=(\sin
k_1^h +\sin k_2^h)/2$.

The energy of the present elementary excitation is,

\begin{eqnarray}
E=-2tL\int \cos k\rho(k)dk+4t{\cal {R}}\sqrt{1-(\Lambda-iu)^2} \, .
\end{eqnarray}
Thus, the corresponding energy deviation in the presence of the external flux can be
expressed as,
\begin{eqnarray}
\Delta E(\phi)=2tL\int \sin k\Delta k\rho(k)dk+ 4t{\cal
{R}}\frac{(\Lambda-iu)\Delta\Lambda}{\sqrt{1-(\Lambda-iu)^2}} \, .
\end{eqnarray}
Note that the momentum of the two charge-distribution holes is determined by the value of
the rapidity $\Lambda$. The important point is that their contributions to the
charge-current cancel in the thermodynamical limit.

Numerical solution of the BA equations with the quantum-number occupancy
configurations given in Eq. (\ref{hub_eq:chargesinglet}) leads to the energy
spectra plotted in Fig. 8. From use of Eq. (\ref{hub_eq:jsjqdef}) we could
confirm that both the spin and charge carried by these elementary excitations
vanish in the thermodynamical limit \cite{new}.
\begin{figure*}
\includegraphics[width=15cm,height=7cm]{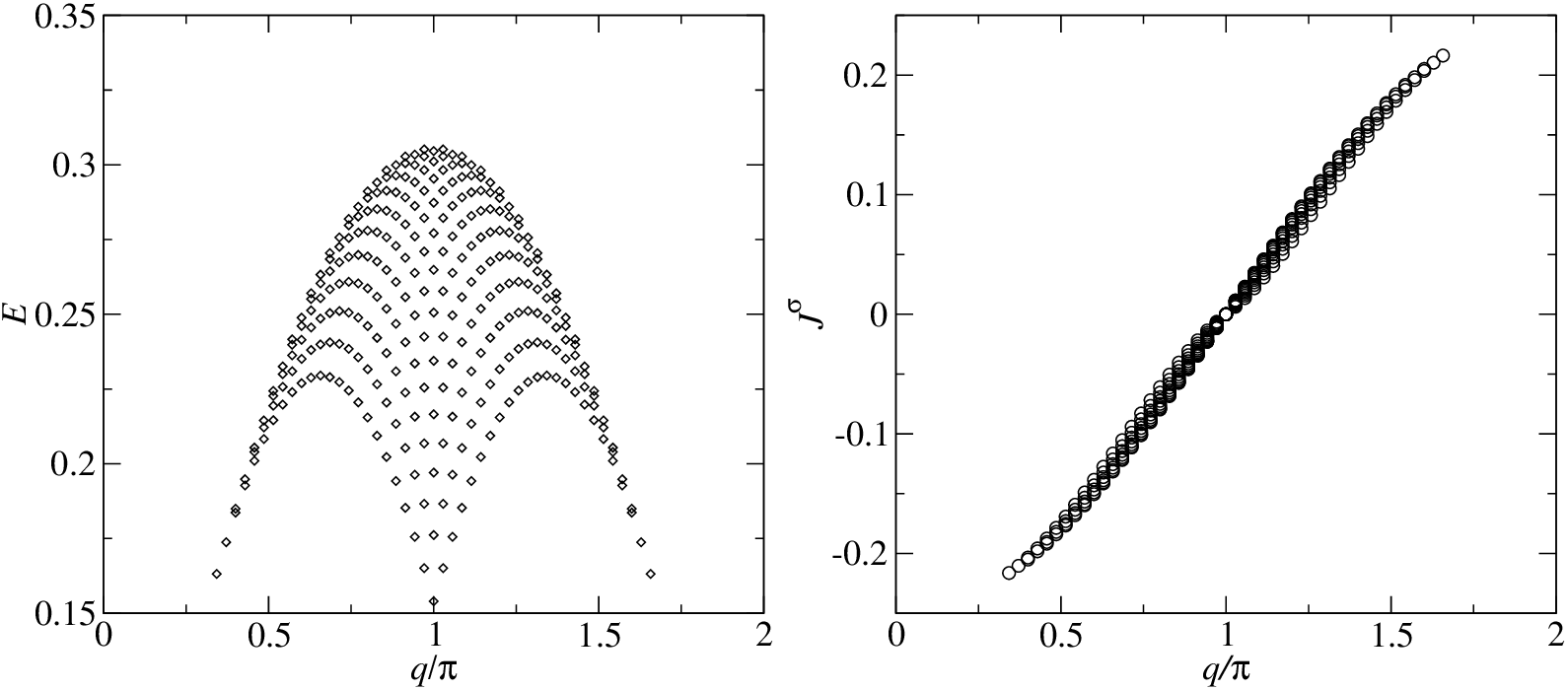}
\caption{\label{fig9} The half-filling energy spectrum ({\bf left}) and
spin-current spectrum ({\bf right}) of the spin excitation with a string of
length two for $u=10$, $N=L=75$,and  $M=25$.\\}
\end{figure*}
\begin{figure*}
\includegraphics[width=15cm,height=7cm]{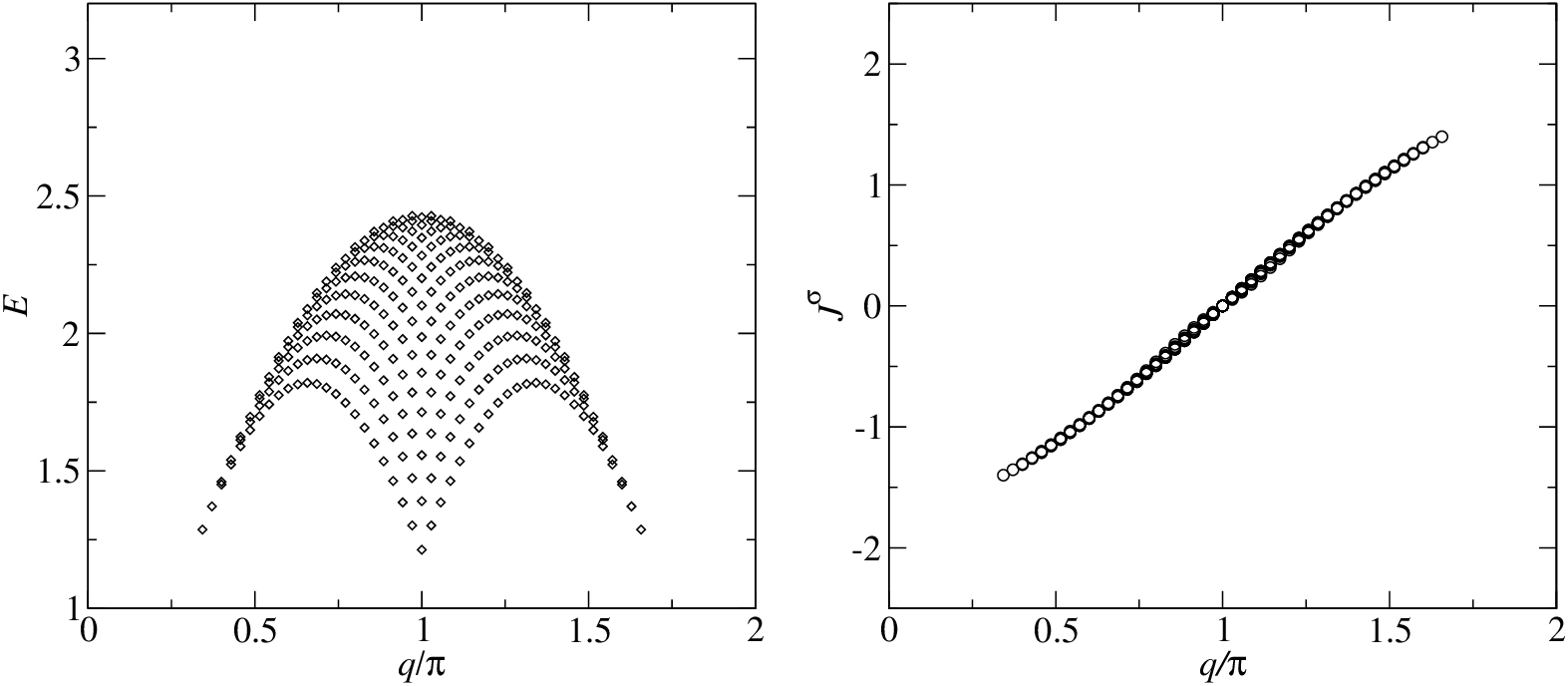}
\caption{\label{fig10} The half-filling energy spectrum ({\bf left}) and
spin-current spectrum ({\bf right}) of the spin excitation with a string of
length two for $u=1$, $N=L=75$,and  $M=25$.\\}
\end{figure*}

\subsection{Half-filling with non-zero
magnetization}\label{hub_sec:fourngg0} \pagestyle{myheadings}

While the energy spectra plotted in figures 1-8 correspond to the general
energy spectrum given in Eq. (\ref{EP}), in the figures 9-14 presented below
the energy spectra of the elementary excitations refer to the general spectrum
provided in Eq. (\ref{E-GS}).

Here we consider that the initial half-filling ground state corresponds to a finite spin
density. That is achieved by the presence of a magnetic field. Thus, for such a ground
state and corresponding elementary excitations one has that $M<N/2$ and the integration
limit of the spin variable $\lambda$ changes from $\infty$ to a finite cut-off. For odd
values of $M$, the distribution of the corresponding spin quantum numbers is still of the
form given in Eq. (\ref{eq:groundstateqn}). The charge excitations do not change much due
to the spin-charge separation. For the spin sector, however, the excitation spectrum
presents qualitative changes.

There are still excitations which increase the value of $S$ by one. However,
such excitations require a minimum finite energy and thus the corresponding
excitation energy spectrum has a gap. The spin distribution of these elementary
excitations displays two holes. Moreover, such elementary excitations carry in
general a finite spin current. Additionally, since the integration limit of
$\lambda$ is now finite, the external flux can shift the whole integration
region. That process also contributes to the spin current. The corresponding
current feature is expected to be a linear function of the momentum $q$. The
same kind of spin-current feature arises now from elementary excitations
involving one spin string of length two. For the case of the zero-magnetization
ground state such excitation is of spin singlet character. However, in contrast
to the zero-spin case now the two spin-distribution holes contribute to the
spin current. Consideration of elementary excitations whose deviations from the
quantum number configuration occupancies is the same as that of equations
(\ref{hub_eq:spinsingletqnc}) for the spin singlet excitation, leads in the
case of an initial ground state with finite spin density to the energy and
spin-current spectra plotted in Figs. 9 and 10 for $u=10$ and $u=1$,
respectively. Note that now such excitations lead indeed to a finite value for
the spin current.

\subsection{Away from half filling}
\label{hub_sec:fourn0a} \pagestyle{myheadings}

Finally, let us consider that the initial ground state is metallic and thus
refers to an electronic density away from half-filling. The ground-state spin
density is considered to be zero. In this case, besides the ground state, there
also exist real rapidity solutions for some of the elementary excitations. For
simplicity, we still consider the case of $N/2$ odd but now with $N<L$. The
simplest elementary excitation away from half-filling corresponds to removing
one $I_j$ from the ground-state occupied charge-distribution sector and adding a
new one outside such a sector. Such an excitation has a ``particle-hole"
character. It is characterized by the following BA numbers,
\begin{eqnarray}
\{I_j\}&=&\left\{-\frac{N-1}{2},\dots, -\frac{N-1}{2}+n-1,\right. \nonumber
\\&& \left. -\frac{N-1}{2}+n+1,\dots ,\frac{N-1}{2}, I_n\right\}.
\end{eqnarray}
where $|I_n|>(N-1)/2$. We consider excitations such that the $\lambda$ spin
distribution remains unchanged.

In figures 11 and 12 we plot the energy and charge-current spectra of such
elementary excitations for $u=10$ and $u=1$, respectively. Note that for strong
coupling ($u=10$) the charge current may have negative values whereas  for weak
coupling ($u=1$) it has always positive values.

In the strong coupling limit the charge BA equation simplifies to,
\begin{eqnarray}
k_j L=2\pi I_j+\phi_\uparrow \, .
\end{eqnarray}
It follows that the energy deviation can be written as,
\begin{eqnarray}
\Delta E(\phi)=2t\sum_j \sin k_j \Delta k_j \, .
\end{eqnarray}
Furthermore, the current reads,
\begin{eqnarray}
J^\rho=2t[\sin(2k_F+q)-\sin(2k_F)] \, , \label{eq:currentofremoveq}
\end{eqnarray}
where $2k_F=\pi N/L$ denotes the ``Fermi momentum" and $q$ the momentum. Note
that Eq. (\ref{eq:currentofremoveq}) provides the charge current for an
excitation where a charge is removed from the "Fermi point" and created outside
the ground-state ``Fermi sea". As long as $2k_F>\pi/2$, the current in the
vicinity of $2k_F$ is negative. On the other hand, the effective charge carried
by this excitation is negative.
\begin{figure*}
\includegraphics[width=15cm,height=7cm]{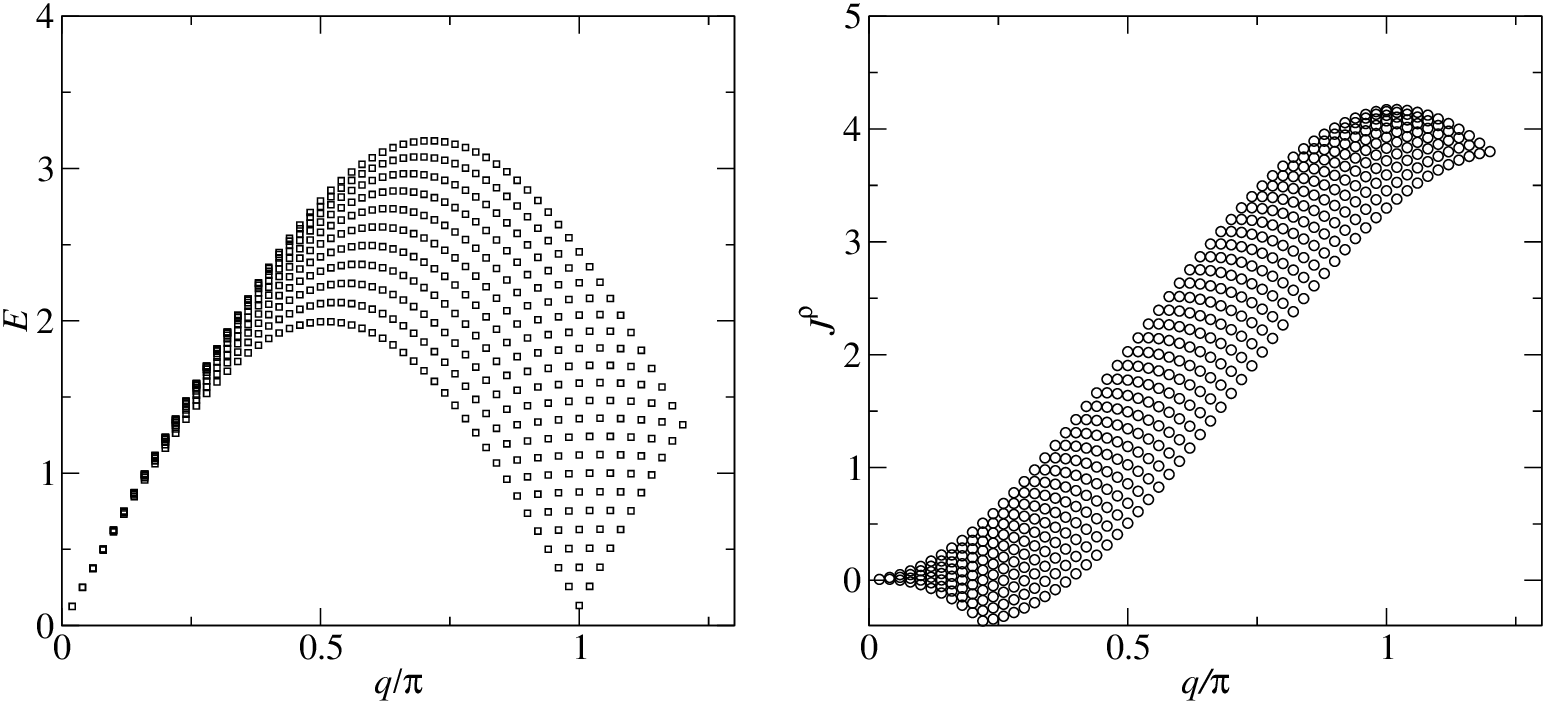}
\caption{\label{fig11} The quarter-filling energy spectrum ({\bf left}) and
charge current spectrum ({\bf right}) of a charge ``particle-hole" excitation
for $u=10$, $L=100, N=50$, and $M=25$.\\\\}
\end{figure*}
\begin{figure*}
\includegraphics[width=15cm,height=7cm]{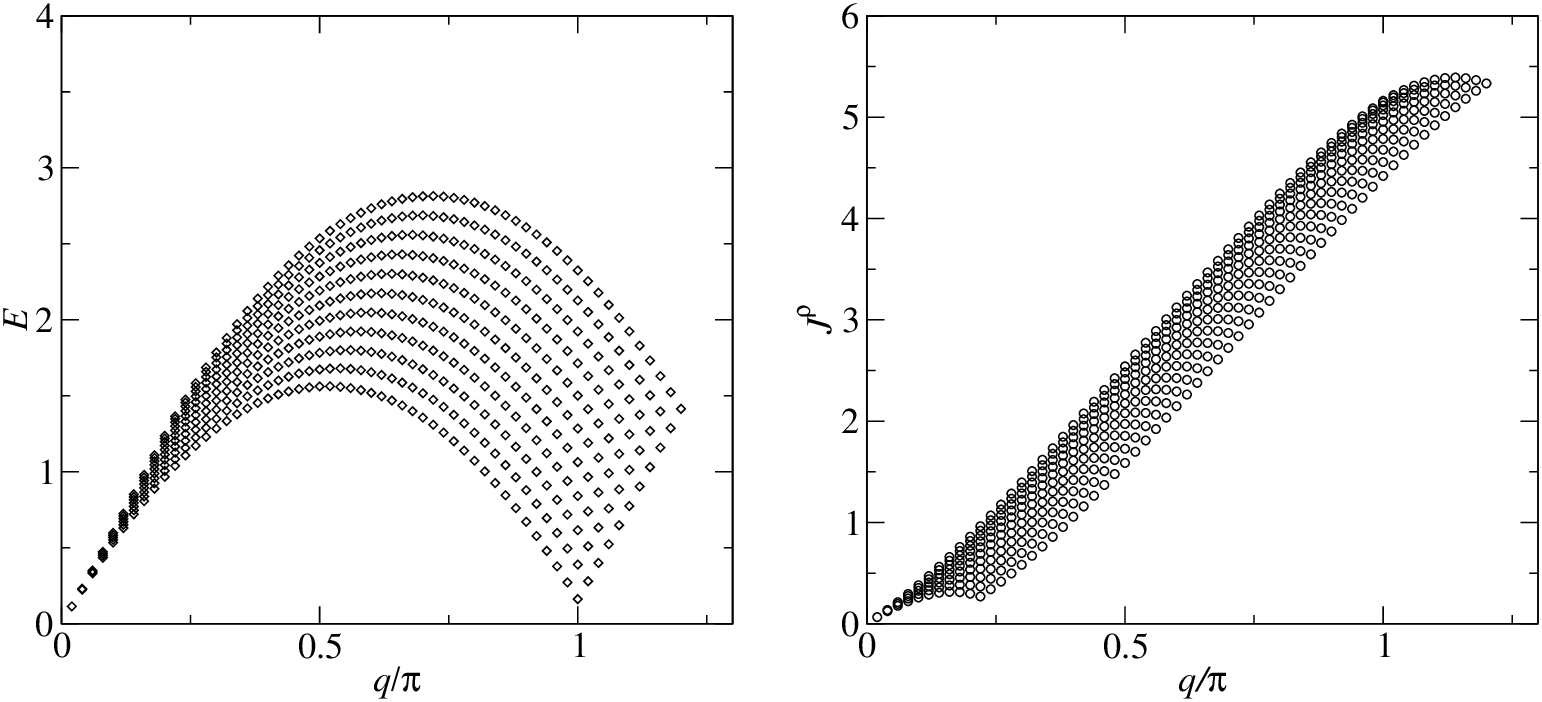}
\caption{\label{fig12} The quarter-filling energy spectrum ({\bf left}) and
charge current spectrum ({\bf right}) of a charge ``particle-hole" excitation
for $u=1$, $L=100, N=50$, and $M=25$.\\}
\end{figure*}

The contribution to the charge current of the charge-distribution hole excitation is
straightforward to obtain and reads,
\begin{eqnarray}
J_h^\rho=2t[\sin 2k_F-\sin k_h],\;\; k_h\in[-2k_F, 2k_F] \, ,
\end{eqnarray}
where $k_h$ is the momentum of the charge-distribution hole. At $2k_F=\pi/2$ it
leads to a feature in the charge current spectrum defined by the function
$[1-\cos q/2]$. Combining the ``particle" and ``hole" contributions one finds,
\begin{eqnarray}
J_p^\rho=2t[\sin(2k_F+k)-\sin k_h] \, .
\end{eqnarray}
The group velocities of the charge ``particle" and ``hole" are such that,
\begin{eqnarray}
v^\rho_p=-2t\sin k_p,\;\;\;\; v^\rho_h=2t\sin k_h \, .
\end{eqnarray}
Hence, the effective charge carried by the ``particle" and the ``hole" are
$e(=-1)$ and $-e(=1)$, respectively. We recall that all this analysis applies to
the strong coupling limit only.

In turn, in the weak coupling limit the interaction between the charge holes is
so weak in the ``Fermi sea" that the charge current is almost a linear function
of the momentum. For the "particle" at quarter filling, however, since the
distribution function $\rho(k)$ becomes a very narrow peak and the ``Fermi
surface" is compressed, the charge current changes from negative to positive.

We have shown above that both the spin and charge currents carried by the
charge $\eta$-spin-singlet excitations containing one charge string of length
one vanish. However, away from half-filling the corresponding charge
excitations containing one charge string of length one have different
properties. The quantum number occupancy configuration of such elementary
excitations is still described by Eq. (\ref{hub_eq:chargesinglet}). We have
solved the corresponding BA equations. The found energy spectrum and
charge-current spectrum are plotted in Figs. 13 and 14 for $u=10$ and $u=1$,
respectively. In contrast to the corresponding excitations relative to the
half-filling ground state, the present excitations carry charge but no spin.
\begin{figure*}
\includegraphics[width=15cm,height=7cm]{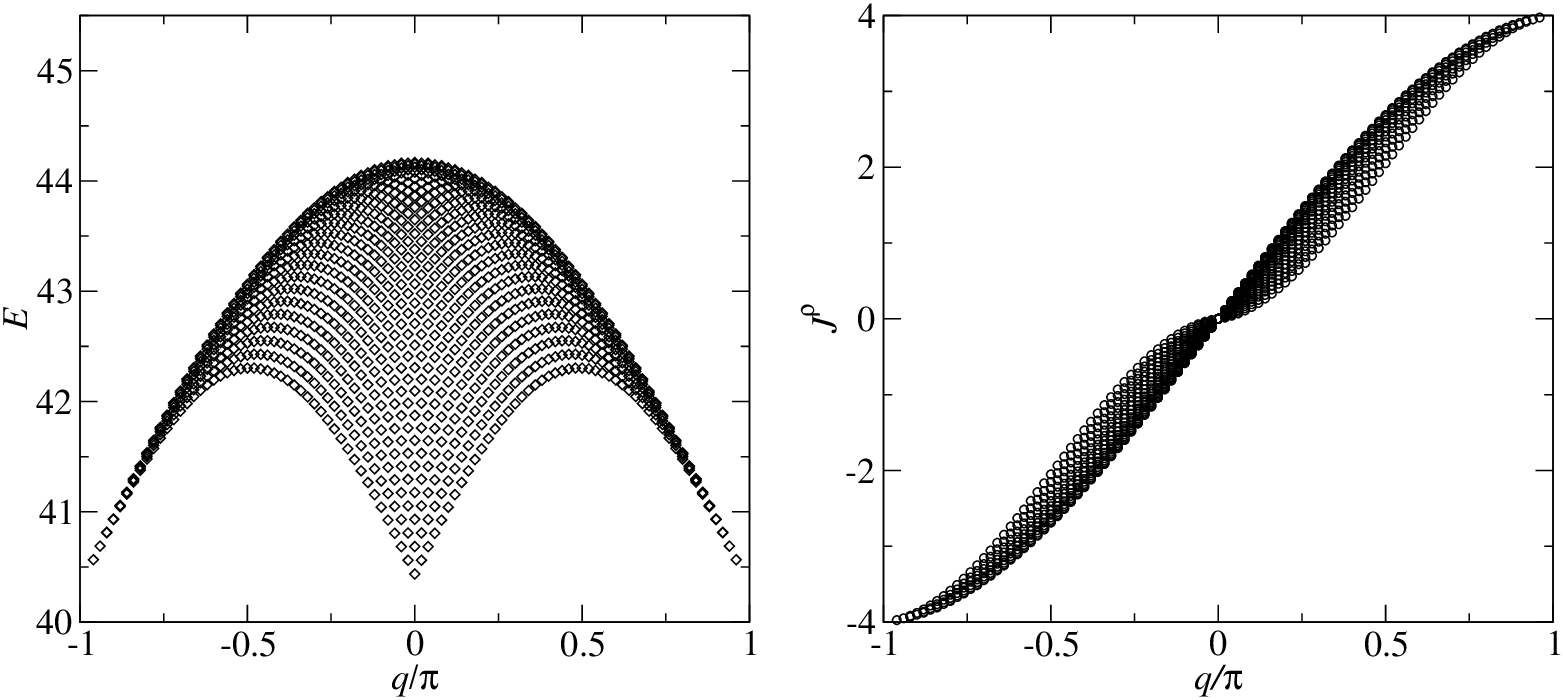}
\caption{\label{fig13} The quarter-filling energy spectrum ({\bf left}) and
charge-current spectrum ({\bf right}) of a charge excitation containing one
charge string of length one at for $u=10$, $L=100, N=50$, and $M=25$.\\}
\end{figure*}
\begin{figure*}
\includegraphics[width=15cm,height=7cm]{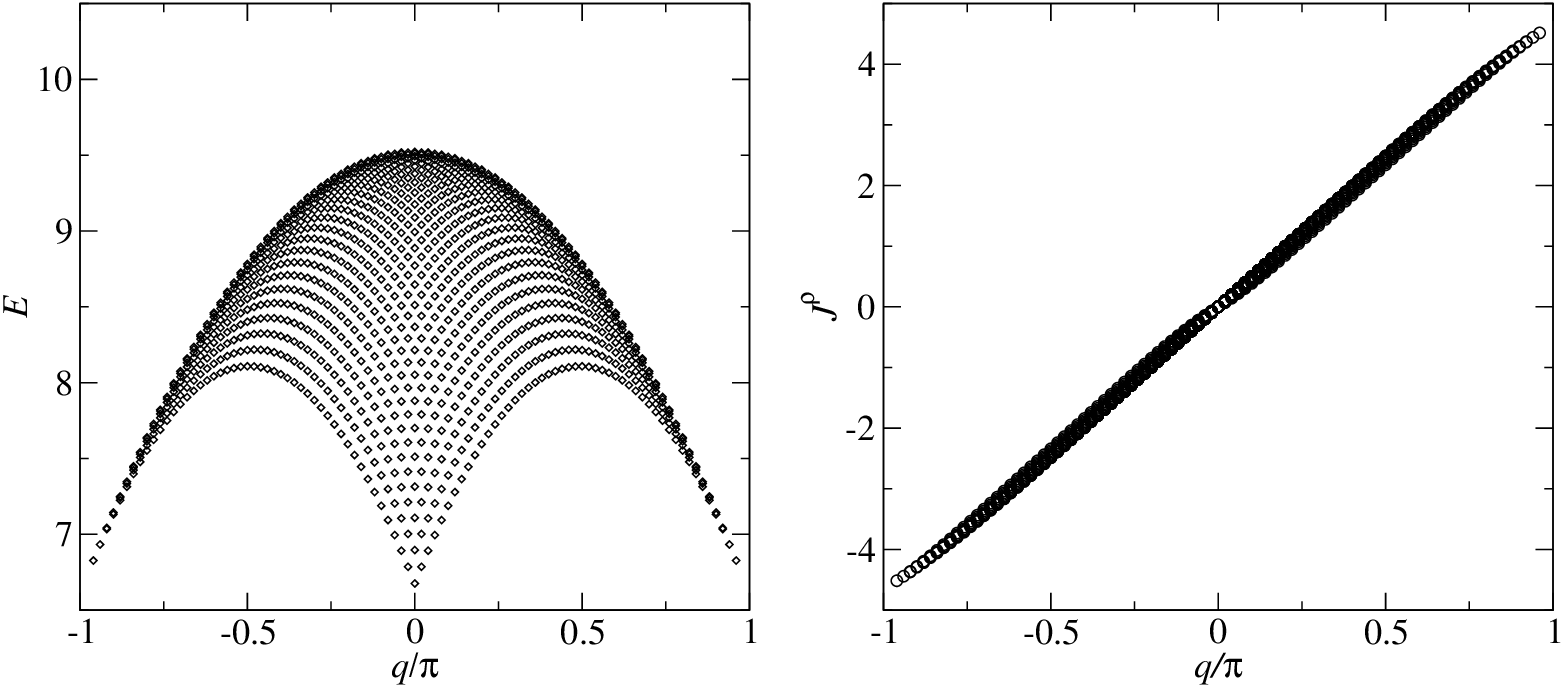}
\caption{\label{fig14} The quarter-filling energy spectrum ({\bf left}) and
charge-current spectrum ({\bf right}) of a charge excitation containing one
charge string of length one at for $u=1$, $L=100, N=50$, and $M=25$.\\}
\end{figure*}

\section{Concluding remarks}

In this paper we have studied the spin and charge currents carried by the elementary
excitations of the 1D Hubbard model. Most of our results refer to half-filling. Both the
charge $\eta$-spin-singlet and spin-singlet elementary excitations considered in our
study carry no charge and no spin. Moreover, the spin-triplet excitations considered in
this paper carry spin but no charge, while the charge $\eta$-spin-triplet elementary
excitations carry charge but no spin.

Our results reveal that the present quantum liquid is not an ideal insulator
for half-filling, in apparent contradiction to the general predictions of Ref.
\cite{Prelovsek}. Indeed, ideal insulating behavior requires that all
half-filling states carry no charge current. However, we note that out of the
three types of charge $\eta$-spin-triplet elementary excitations of $\eta$-spin
projections $0,\pm1$, the states with projection $0$ are indeed half-filling
states. Our results show that such charge $\eta$-spin-triplet states carry
finite charge current both in the strong and weak coupling limits. According to
the results of Ref. \cite{Prelovsek}, this implies a finite value for the
charge stiffness $D (T)$ at finite temperatures $T>0$. This result agrees
qualitatively with the studies of Ref. \cite{Kawa}, which lead to a finite
value for $D(T)$ at half filling and $T>0$. However, such studies did not take
into account the $\eta>0$ and $\eta_z=0$ states which carry the charge current at
half filling and lead to $D(T)>0$ for $T>0$. Moreover, both the studies
of the present paper and the related studies of
Ref. \cite{new} provide strong evidence that the $\eta=0$ and $\eta_z =0$ states
considered in Ref. \cite{Kawa} do not carry charge current and do not
contribute to $D(T)$.


The model studied here is integrable and according to the general
arguments of Ref. \cite{Prelovsek} should display ideal insulating behavior at
half filling. That issue is clarified elsewhere by study of the  microscopic mechanism
which is behind the half-filling properties concerning transport of charge \cite{new}.

Finally, our results reveal the occurrence of charge-spin separation at finite energies,
since some of the elementary excitations studied here have an energy gap relative to the
initial ground state. Such a finite-energy charge-spin separation deserves further
studies. We note that the photoemission studies of Ref. \cite{Sing} have detected a
spin-charge separation in quasi-1D compounds for the whole energy bandwidth.

\section*{Acknowledgments} S. J. G. thanks the support of the
University of \'Evora where this research was initiated and acknowledges
the financial support of Earmarked Grant for Research from the Research Grants
Council of HKSAR, China (Project No CUHK 400906). J. M. P. C. and N. M. R. P.
thank the support of the ESF Science Program INSTANS
and grants POCTI/FIS/58133/2004 and PTDC/FIS/64926/2006.

\end{document}